\pdfoutput=1

\documentclass[11pt]{article}

\usepackage[]{ACL2023}

\usepackage{times}
\usepackage{latexsym}

\usepackage{booktabs} %
\usepackage{amsmath}
\usepackage{amsfonts}
\usepackage{comment}
\usepackage{footmisc}
\usepackage{mathrsfs}
\usepackage{graphicx}
\usepackage{subfigure}
\usepackage{multirow}
\usepackage{algorithm}
\usepackage{algorithmic}
\usepackage{multicol}  
\usepackage{enumitem}
\usepackage{array}
\usepackage{float}
\usepackage{hyperref}
\usepackage{ulem}
\usepackage{color, xspace}
\usepackage{makecell}
\usepackage[T1]{fontenc}

\usepackage[utf8]{inputenc}

\usepackage{microtype}

\usepackage{inconsolata}

\definecolor{greyblue}{rgb}{0.188, 0.463, 0.702}
\definecolor{CamelliaRed}{rgb}{0.93, 0.247, 0.302}
\definecolor{Youhuang}{rgb}{0.987,0.524,0.151}
\definecolor{Watergreen}{rgb}{0.494,0.086,0.443}
\usepackage{color, xspace}

\title{MILL: Mutual Verification with Large Language Models for Zero-Shot Query Expansion}

\author{Pengyue Jia$^1$, Yiding Liu$^2$, Xiangyu Zhao$^1$\thanks{*Corresponding author}, Xiaopeng Li$^2$\\
        \textbf{Changying Hao}$^2$, \textbf{Shuaiqiang Wang}$^2$, \textbf{Dawei Yin}$^2$ \\
        $^1$City University of Hong Kong, $^2$Baidu Inc. \\
        \texttt{\{jia.pengyue,xiaopli2-c\}@my.cityu.edu.hk} \\
        \texttt{\{liuyiding.tanh,cyhaocn,shqiang.wang\}@gmail.com} \\
        \texttt{{xianzhao}@cityu.edu.hk}, \texttt{{yindawei}@acm.org}
        }

\begin{document}
\maketitle
\begin{abstract}
  Query expansion, pivotal in search engines, enhances the representation of user information needs with additional terms. 
  While existing methods expand queries using retrieved or generated contextual documents, each approach has notable limitations.
  Retrieval-based methods often fail to accurately capture search intent, particularly with brief or ambiguous queries.	
  Generation-based methods, utilizing large language models (LLMs), generally lack corpus-specific knowledge and entail high fine-tuning costs. 
  To address these gaps, we propose a novel zero-shot query expansion framework utilizing LLMs for mutual verification. 
  Specifically, we first design a query-query-document generation method, leveraging LLMs' zero-shot reasoning ability to produce diverse sub-queries and corresponding documents. 
  Then, a mutual verification process synergizes generated and retrieved documents for optimal expansion. 
  Our proposed method is fully zero-shot, and extensive experiments on three public benchmark datasets are conducted to demonstrate its effectiveness over existing methods. 
  Our code is available online at \href{https://github.com/Applied-Machine-Learning-Lab/MILL}{https://github.com/Applied-Machine-Learning-Lab/MILL} to ease reproduction.
\end{abstract}

\section{Introduction}

Query expansion is a critical technique in search systems, aiming to effectively capture and represent users' information needs~\cite{zhu2023large,efthimiadis1996query}. 
Search engines employ query expansion to resolve ambiguities in queries and align the vocabulary of queries and documents. 
Central to this task is the development of contextual documents, comprising additional query terms, to enhance effectiveness~\cite{azad2019query}.

Specifically, existing research predominantly falls into two categories: retrieval-based and generation-based methods. 
Retrieval-based methods~\cite{lv2010positional,yan2003multimedia,li2022improving} typically construct contextual documents from the targeted corpus, assuming that the top-retrieved documents (i.e., pseudo-relevance feedback (PRF)) are reasonable expansions of a given query. 
Generation-based methods~\cite{jagerman2023query,mao2023large,wang2023query2doc} often utilize advanced generative models, such as Large Language Models, as an external knowledge base for producing contextual documents.

However, both methods have clear limitations.
For retrieval-based methods, it has been observed in practice that the documents retrieved with the original query do not align well with the information needs, particularly when the original query itself is brief and ambiguous \cite{cao2008selecting,jagerman2023query}. 
For generation-based methods, directly using off-the-shelf LLMs in a few-shot or zero-shot manner can hardly align the model with a specific corpus~\cite{wang2023query2doc}. 
In contrast, the LLMs could easily generate useless out-of-domain information.

To this end, we propose a novel query expansion framework based on Large Language Models (LLMs), integrating both retrieved and generated documents to mitigate their respective limitations.
First, to improve contextual document generation, we design a query-query-document prompt that leverages an LLM as a zero-shot reasoner to decompose a query into multiple sub-queries during contextual document generation. 
This helps the LLM generate diverse contextual information that is more likely to cover the underlying search intent.

\begin{figure*}[!ht]
    \centering
    \includegraphics[width=\textwidth]{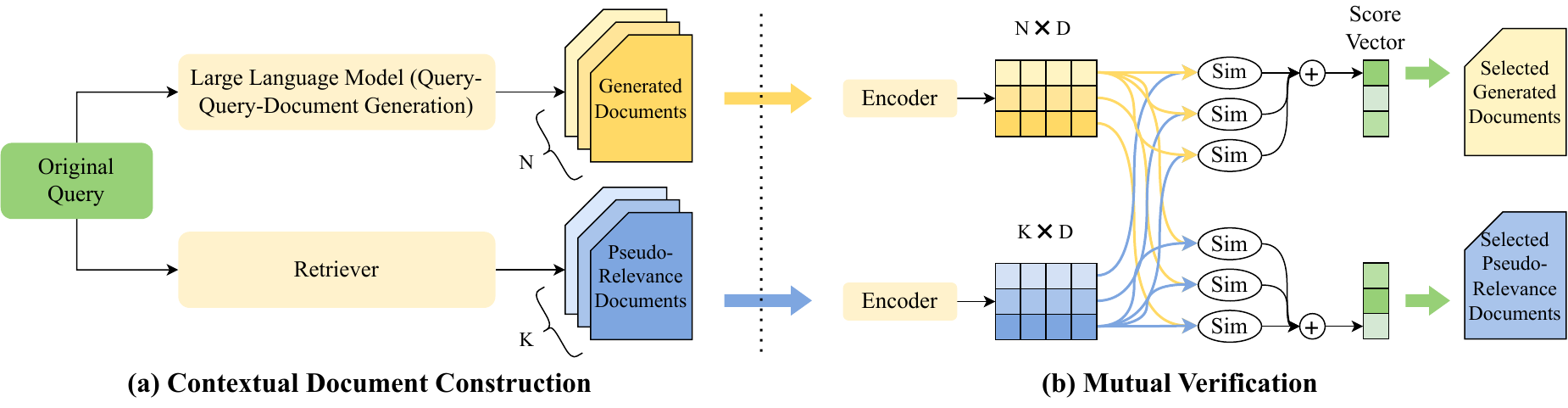}
    \caption{Overview of MILL. }
    \label{fig:overview}
\end{figure*}

Next, we propose a mutual verification framework that exploits generated and retrieved contextual documents for query expansion. 
To be more specific, we propose to filter out the uninformative generated documents via comparing their relevance with the top-retrieved documents. 
By doing this, the selected generated documents are intuitively more aligned with the target corpus. 
Conversely, we also filter out the noisy retrieved documents via comparing their relevance with the generated documents. 
The external contextual knowledge embedded in the generated documents can facilitate the retrieved documents to more accurately reveal search intent.
We evaluate the proposed method on the downstream information retrieval task in a zero-shot manner. 
The results on three public datasets demonstrate that our proposed method significantly outperforms the state-of-the-art baselines. 
Overall, the contributions can be summarized as follows:
\begin{itemize}[leftmargin=*]
    \item We propose a \textbf{M}utual Ver\textbf{I}fication method with \textbf{L}arge \textbf{L}anguage model (denoted as MILL), a novel framework that combines generated and retrieved context for query expansion. MILL is able to 
    mitigate the limitations of generated and retrieved context, 
    and thus can provide more high-quality context for query expansion.
    \item To improve the generated contextual documents, 
    we design a query-query-document prompting method, which elicits richer and more diverse knowledge from LLMs to cover the underlying search intents and information needs of users.
    \item MILL can perform high-quality query expansion in a zero-shot manner. 
    We conduct extensive experiments on the downstream information retrieval task on three public datasets. The results demonstrate that MILL significantly outperforms existing retrieval and generation-based methods.
\end{itemize}

\section{Problem Definition}

Given a user query $q$, query expansion is to apply a function $f$ to expand $q$ with additional contextual information: $q^{\prime} = f_\theta(q)$, where $\theta$ represents the parameters.
Using the expanded query $q^{\prime}$ should be able to achieve better downstream retrieval performance compared to the original query $q$. More formally, such an objective can be defined as
\begin{equation}
    \underset{\theta}{\operatorname{argmax}} \ \mathcal{M}(q^\prime, R), \ \text{where}\ q^{\prime} = f_\theta(q).
\end{equation}
where $\mathcal{M}$ denotes the evaluation metric of the retrieval performance (e.g., recall, NDCG), and $R$ denotes the retrieval model.

\section{Methodology}

In this section, we introduce our proposed query expansion method in detail. Specifically, we give an overview of MILL in Section \ref{overview}, elaborate the query-query-document generation in Section \ref{qqd}, and introduce the mutual verification framework in Section \ref{mv}.

\begin{figure*}[!ht]
    \centering
    \includegraphics[width=\textwidth]{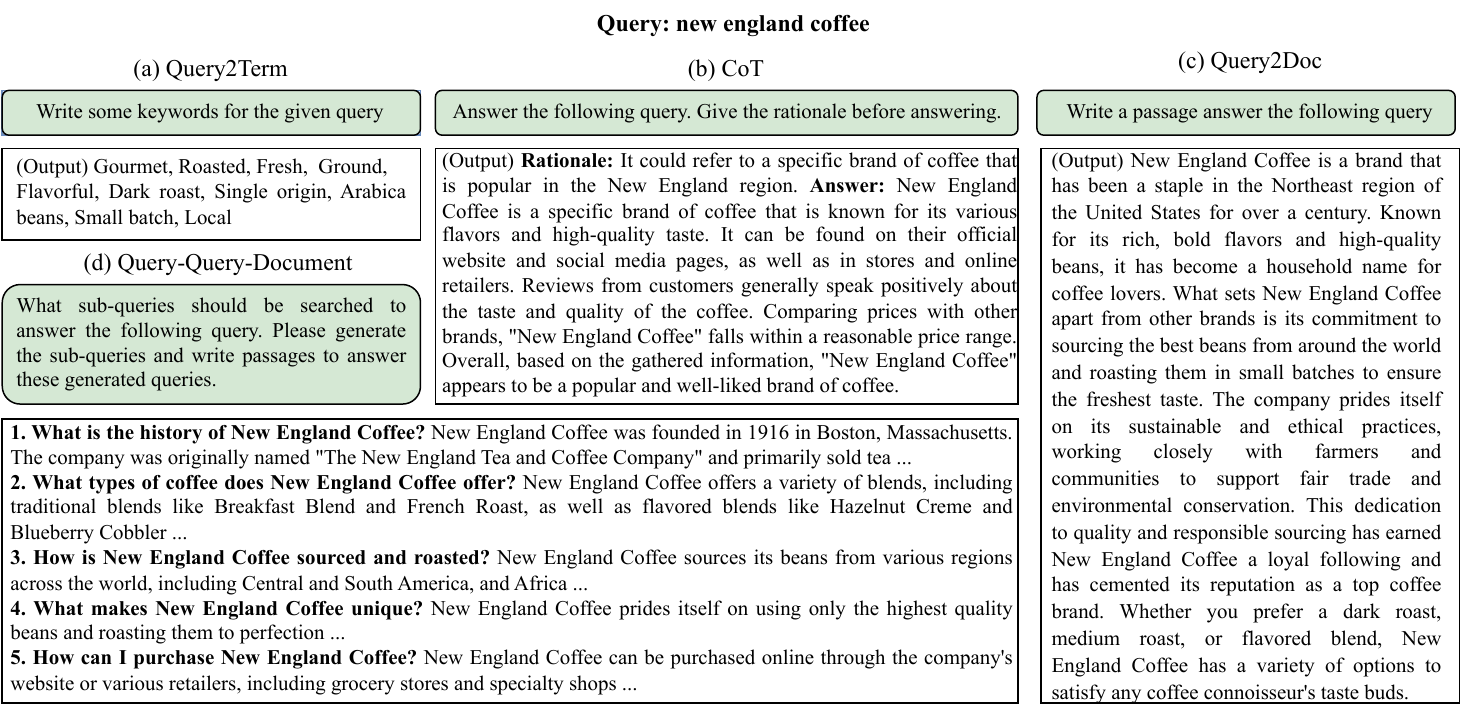}
    \caption{Query-query-document prompt compared to Query2Term, CoT, and Query2Doc. Query-query-document instructs the LLM to expand the original query from multiple perspectives by inferring the sub-queries and generating corresponding contextual documents.}
    \label{fig:prompt_framework}
\end{figure*}

\subsection{Overview} \label{overview}

The overall workflow of MILL is depicted in Figure \ref{fig:overview}, which comprises two steps, i.e., contextual document construction (Figure \ref{fig:overview}(a)) and mutual verification (Figure \ref{fig:overview}(b)). In particular, the two steps focus on the \emph{diversity} and \emph{quality} of contextual documents, respectively.

In the \textbf{contextual document construction} stage, we aim to construct diverse sets of contextual documents via both retrieval and generation. 
To create diverse contextual documents via generation, we propose a query-query-document prompt, which instructs an LLM to generate sub-queries and contextual documents in a step-by-step manner. This can better leverage the reasoning ability of LLMs that decompose a given query expansion task into multiple sub-tasks, where the generated documents could be more diverse and informative.

In the \textbf{mutual verification} stage, we aim to identify those high-quality contextual documents constructed in the first stage. 
In particular, the mutual verification leverages the strengths of generated documents in implying the search intent and the domain-specific nature of PRF documents, enabling a reciprocal selection between the two types of contextual documents. As a result, the finalized documents are more high-quality query expansion to be applied in downstream retrieval tasks.

\subsection{Query-Query-Document Generation} \label{qqd}

Recently, a handful of studies \cite{wang2023query2doc, jagerman2023query} have explored using Large Language Models (LLMs) to expand queries and gain initial success. However, most of them use a rather simple prompt for document generation, e.g., ``write a passage that answers the given query''. For a brief or ambiguous query that has multiple possible intents, the generation results could easily miss the real search intent. Motivated by this, we design a novel zero-shot prompt, particularly for the query expansion task. 
This method can exploit the reasoning ability of LLMs to first decompose the original query into multiple sub-queries before document generation. This improves generation diversity, and the contextual documents are more likely to cover the real search intent.

As shown in Figure \ref{fig:prompt_framework}(d), we use the instruction "what sub-queries should be searched to answer the following query: \{query\}." to generate sub-queries that further clarify the original query.
At the same time, we instruct the language model to generate contextual documents for each sub-query through "Please generate the sub-queries and write passages to answer these generated queries." By doing this, we finally have multiple sub-queries and their corresponding contextual documents, which are more likely to cover the user's search intent. Note that the proposed method is zero-shot, which can be easily extended to few-shot.

\subsection{Mutual Verification} \label{mv}

Next, we elaborate on the mutual verification framework,
where we leverage the aforementioned generated documents and pseudo-relevance documents (i.e., the retrieval-based contextual documents) to improve the overall quality of query expansion. The intuition is to leverage two types of information to complement each other, which are 1) the corpus-specific domain information of retrieved pseudo-relevance documents, and 2) the generated information of LLM reasoning that is more likely to uncover real search intent.

More specifically, the inputs of mutual verification have two sets of contextual documents:
\begin{gather}
    \mathcal{D}^{\text{LLM}} = \{ d_{n}^{\text{LLM}} \} = \text{LLM}(p,q),\  n \in ( 0,N ] \\
    \mathcal{D}^{\text{PRF}} = \{ d_{k}^{\text{PRF}} \} = R_r(q),\  k \in (0,K]
\end{gather}
where $\mathcal{D}^{\text{LLM}}$ represents the $N$ LLM-generated documents with query-query-document prompt (denoted as $p$), and $\mathcal{D}^{\text{PRF}}$ represents the $K$ documents retrieved by a vanilla PRF method (denoted as $R_r$), e.g., BM25 retrieval. 
Note that each generated document comprises a series of sub-queries and their corresponding passages.

Next, we aim to rerank the documents in $\mathcal{D}^{\text{LLM}}$ and $\mathcal{D}^{\text{PRF}}$. In specific, we first use an off-the-shelf dense representation model to compute the representation (i.e., $\mathbf{x}_n^\text{LLM}$ or $\mathbf{x}_k^\text{PRF}$) of each document (i.e., $d_n^\text{LLM}$ or $d_k^\text{PRF}$) as
\begin{gather}
    \mathbf{x}^{\text{LLM}}_{n} = \textrm{Encoder}(d_{n}^{\text{LLM}}) \label{answer_encode}, \\
    \mathbf{x}^{\text{PRF}}_{k} = \textrm{Encoder}(d_{k}^{\text{PRF}}) \label{document_encode},
\end{gather}
where $\mathbf{x}^{\text{LLM}}_{n}$ denotes the vector for $n$-th generated document and $\mathbf{x}^{\text{PRF}}_{k}$ denotes the vector for $k$-th pseudo-relevance documents.\

Then, we compute the semantic relevance between every pair of $d_n$ and $d_k$ with cosine similarity (denoted as $\text{sim}(\cdot)$), and assign a score to every document as
\begin{gather}
    s_n^{\text{LLM}} = \sum\nolimits_{k=1}^{K} \textrm{sim}(\mathbf{x}_n^{\text{LLM}}, \mathbf{x}^{\text{PRF}}_{k}), \\
    s_k^{\text{PRF}} = \sum\nolimits_{n=1}^{N} \textrm{sim}(\mathbf{x}_k^{\text{PRF}}, \mathbf{x}_n^{\text{LLM}}). 
\end{gather}
Here, we score every generated document $d_n^\text{LLM}$ via aggregating its semantic relevance scores with all pseudo-relevance documents. Therefore, the score $s_n^\text{LLM}$ can be interpreted as how well $d_n^\text{LLM}$ is aligned with the target corpus. On the other hand, the score $s_k^\text{PRF}$ can be viewed as how well the retrieved document $d_k^\text{PRF}$ is likely to be a reasonable context judged by the reasoning results of LLM.

Finally, we select the top-scored documents in both sets as the final contextual documents as
\begin{equation}
\resizebox{\linewidth}{!}{$
    \begin{aligned}
            \mathcal{D}^{\text{LLM}}_{s} &= \{d_{n}^{\text{LLM}}\}, \  n \in \{ n \  | s_{n}^{\text{LLM}} \in TopN^{\prime}(s^{\text{LLM}}) \}, \\
        \mathcal{D}^{\text{PRF}}_{s} &= \{d_{k}^{\text{PRF}}\}, \ k \in \{ k \ | s_{k}^{\text{PRF}} \in TopK^{\prime}(s^{\text{PRF}}) \},
    \end{aligned}
$}
\end{equation}
where $\mathcal{D}^{\text{LLM}}_{s}$ and  $\mathcal{D}^{\text{PRF}}_{s}$ are the final selected document sets.

\subsection{Query Expansion for Retrieval}
After mutual verification, we integrate the selected generated documents and pseudo-relevance documents with the original query to perform the final retrieval task. In particular, we concatenate them as the new query $q^\prime$ as: 
\begin{gather}
    q^{\prime} = \textrm{concat}(q,\ q,\ q, \ q,\ q,\ \mathcal{D}_{s}^{\text{PRF}},\ \mathcal{D}_{s}^{\text{LLM}})
\end{gather}
We repeat the original query 5 times following papers~\cite{wang2023query2doc,jagerman2023query} to emphasize its significance. It is worth noting that the proposed query expansion method does not need any additional labeled data and model fine-tuning. Such a zero-shot method with off-the-shelf LLM and retriever has huge potential to be applied in various search systems.

\section{Experiments}

\begin{table*}[!h]
\caption{Overall comparison on TREC-DL-2019 and TREC-DL-2020. The optimal results are highlighted in bold, while the suboptimal results are underscored. The results are reported on NDCG@N, AP@N, Recall@N, and MRR@N with $\text{N}\in\{10, 100, 1000\}$. The improvements are all significant (i.e., two-sided t-test with $p<0.05$) between the optimal and suboptimal results.}
\centering
\label{tab:overall_result}
\resizebox{\textwidth}{!}{
\begin{tabular}{cccccccccccccc} 
\toprule
\multicolumn{2}{c}{\multirow{2}{*}{Metrics}}                   & \multicolumn{3}{c}{NDCG}                         & \multicolumn{3}{c}{AP}                           & \multicolumn{3}{c}{Recall}                       & \multicolumn{3}{c}{MRR}                           \\ 
\cmidrule{3-14}
\multicolumn{2}{c}{}                                           & @10            & @100           & @1000          & @10            & @100           & @1000          & @10            & @100           & @1000          & @10            & @100           & @1000           \\ 
\midrule
\multirow{25}{*}{TREC-DL-2019} & No expansion                  & 47.95          & 48.74          & 59.34          & 10.14          & 29.07          & 37.00          & 12.23          & 44.22          & 73.62          & 79.44          & 79.49          & 79.50           \\ 
\cmidrule{2-14}
                               & Traditional expansion methods &                &                &                &                &                &                &                &                &                &                &                &                 \\
                               & Bo1                           & 50.86          & 50.01          & 61.09          & 10.98          & 31.08          & 39.99          & 12.85          & 45.42          & 75.11          & 78.75          & 78.81          & 78.81           \\
                               & KL                            & 50.57          & 49.84          & 60.82          & 10.95          & 30.94          & 39.77          & 12.84          & 45.26          & 74.66          & 78.44          & 78.50          & 78.50           \\
                               & RM3                           & 51.56          & 50.41          & 61.23          & 10.78          & 31.70          & 40.45          & 13.14          & 46.37          & 75.43          & 78.94          & 79.01          & 79.01           \\
                               & AxiomaticQE                   & 47.95          & 48.74          & 59.34          & 10.14          & 29.07          & 37.00          & 12.23          & 44.22          & 73.62          & 79.44          & 79.49          & 79.50           \\ 
\cmidrule{2-14}
                               & LLM-based expansion methods   &                &                &                &                &                &                &                &                &                &                &                &                 \\
                               & Query2Term                    & 44.17          & 42.95          & 55.21          & 9.08           & 22.61          & 29.91          & 11.06          & 37.10          & 68.81          & 71.61          & 71.77          & 71.77           \\
                               & Query2Term-FS                 & 50.38          & 49.54          & 61.67          & 11.30          & 29.40          & 37.51          & 12.52          & 43.83          & 76.13          & 75.22          & 75.73          & 75.73           \\
                               & Query2Term-PRF                & 48.56          & 48.08          & 57.63          & 11.05          & 30.78          & 37.18          & 12.24          & 43.69          & 70.20          & 80.10          & 80.17          & 80.19           \\
                               & Query2Doc                     & 62.77          & 61.45          & 71.75          & 13.68          & 39.28          & 49.04          & 14.78          & 52.25          & 84.21          & 90.89          & 91.04          & 91.04           \\
                               & Query2Doc-FS                  & \textbf{63.83} & 61.42          & 72.02          & \uline{14.30}  & 39.54          & 49.65          & 15.44          & 52.57          & 83.75          & 90.55          & 90.55          & 90.56           \\
                               & Query2Doc-PRF                 & 59.00          & 57.47          & 68.23          & 12.15          & 35.29          & 44.56          & 14.39          & 50.29          & 82.12          & 86.63          & 86.63          & 86.63           \\
                               & CoT                           & 63.44          & 59.57          & 70.94          & 13.43          & 35.53          & 45.67          & 14.97          & 49.91          & 83.43          & \textbf{92.61} & \textbf{92.61} & \textbf{92.61}  \\
                               & CoT-PRF                       & 61.63          & 56.81          & 67.85          & 13.13          & 34.84          & 44.73          & 14.82          & 49.02          & 80.37          & 91.47          & 91.72          & 91.72           \\ 
\cmidrule{2-14}
                               & Ensembled expansion methods    &                &                &                &                &                &                &                &                &                &                &                &                 \\
                               & Query2Term$^*$                & 57.26          & 55.93          & 67.18          & 13.07          & 36.18          & 45.48          & 14.60          & 50.12          & 81.13          & 83.53          & 83.86          & 83.86           \\
                               & Query2Term-FS$^*$             & 54.16          & 54.46          & 65.14          & 12.38          & 35.76          & 44.74          & 14.08          & 49.58          & 79.03          & 78.88          & 79.10          & 79.11           \\
                               & Query2Term-PRF$^*$            & 52.17          & 51.84          & 61.49          & 11.94          & 33.93          & 41.94          & 13.69          & 47.26          & 74.36          & 79.07          & 79.25          & 79.25           \\
                               & Query2Doc$^*$                 & 63.59          & 61.74          & 72.41          & 13.98          & 40.81          & 51.31          & 15.37          & 53.94          & 84.78          & 91.28          & 91.49          & 91.49           \\
                               & Query2Doc-FS$^*$              & 64.05          & \uline{62.10}  & \uline{72.79}  & 13.88          & 41.02          & \uline{51.55}  & 15.47          & 54.23          & 84.90          & \uline{92.29}  & \uline{92.29}  & \uline{92.29}   \\
                               & Query2Doc-PRF$^*$             & 62.34          & 61.78          & 72.35          & 13.84          & \uline{41.22}  & 51.44          & 15.19          & \textbf{54.28} & \uline{85.44}  & 89.05          & 89.24          & 89.24           \\
                               & CoT$^*$                       & 64.77          & 61.30          & 72.08          & 14.05          & 39.08          & 49.56          & \uline{15.73}  & 52.35          & 84.23          & 92.19          & 92.19          & 92.19           \\
                               & CoT-PRF$^*$                   & 56.37          & 55.05          & 65.42          & 12.78          & 36.11          & 45.07          & 14.63          & 49.49          & 78.08          & 82.56          & 82.93          & 82.93           \\ 
\cmidrule{2-14}
                               & MILL                          & \uline{63.80}  & \textbf{62.50} & \textbf{73.74} & \textbf{14.75} & \textbf{41.96} & \textbf{53.11} & \textbf{16.17} & \uline{54.26}  & \textbf{85.92} & 91.69          & 91.81          & 91.81           \\ 
\bottomrule
\toprule
\multirow{25}{*}{TREC-DL-2020} & No expansion                  & 49.36          & 50.26          & 59.81          & 14.27          & 31.42          & 35.87          & 17.61          & 50.47          & 75.12          & 80.21          & 80.21          & 80.21           \\ 
\cmidrule{2-14}
                               & Traditional expansion methods &                &                &                &                &                &                &                &                &                &                &                &                 \\
                               & Bo1                           & 49.47          & 53.25          & 63.11          & 14.79          & 34.43          & 39.67          & 17.74          & 54.66          & 79.48          & 80.83          & 80.99          & 80.99           \\
                               & KL                            & 49.27          & 53.20          & 63.01          & 14.68          & 34.31          & 39.53          & 17.66          & 54.70          & 79.39          & 80.83          & 80.99          & 80.99           \\
                               & RM3                           & 50.43          & 54.02          & 63.47          & 14.93          & 35.13          & 40.22          & 17.89          & 55.80          & 79.94          & 78.49          & 78.59          & 78.59           \\
                               & AxiomaticQE                   & 49.36          & 50.26          & 59.81          & 14.27          & 31.42          & 35.87          & 17.61          & 50.47          & 75.12          & 80.21          & 80.21          & 80.21           \\ 
\cmidrule{2-14}
                               & LLM-based expansion methods   &                &                &                &                &                &                &                &                &                &                &                &                 \\
                               & Query2Term                    & 50.12          & 52.43          & 62.27          & 13.12          & 33.06          & 38.49          & 17.39          & 54.61          & 79.07          & 78.74          & 78.77          & 78.78           \\
                               & Query2Term-FS                 & 47.80          & 49.16          & 60.50          & 13.33          & 30.16          & 35.59          & 15.82          & 50.22          & 78.76          & 79.38          & 79.83          & 79.83           \\
                               & Query2Term-PRF                & 47.76          & 48.92          & 59.57          & 12.32          & 29.03          & 33.70          & 14.70          & 49.29          & 76.68          & 78.97          & 79.29          & 79.29           \\
                               & Query2Doc                     & 61.22          & 60.13          & 69.97          & \uline{19.06}  & 41.31          & 47.03          & 21.57          & 57.58          & 83.38          & 88.27          & 88.44          & 88.44           \\
                               & Query2Doc-FS                  & 61.45          & 59.30          & 69.40          & 18.94          & 39.75          & 45.27          & \uline{21.65}  & 56.30          & 82.57          & 90.32          & 90.37          & 90.38           \\
                               & Query2Doc-PRF                 & 55.28          & 57.60          & 67.09          & 17.00          & 38.21          & 43.49          & 19.74          & 58.50          & 82.57          & 84.22          & 84.49          & 84.49           \\
                               & CoT                           & 58.39          & 56.74          & 67.02          & 18.15          & 37.32          & 42.34          & 21.51          & 54.02          & 80.11          & 88.02          & 88.02          & 88.03           \\
                               & CoT-PRF                       & 60.81          & 58.41          & 67.47          & 19.02          & 39.27          & 44.04          & 21.71          & 56.84          & 80.49          & 89.00          & 89.00          & 89.00           \\ 
\cmidrule{2-14}
                               & Ensembled expansion methods    &                &                &                &                &                &                &                &                &                &                &                &                 \\
                               & Query2Term$^*$                & 53.17          & 55.14          & 65.08          & 14.53          & 36.30          & 41.51          & 18.07          & 56.79          & 81.61          & 83.89          & 84.10          & 84.10           \\
                               & Query2Term-FS$^*$             & 50.95          & 52.11          & 62.42          & 13.80          & 33.47          & 38.34          & 16.98          & 53.55          & 79.38          & 81.77          & 81.81          & 81.81           \\
                               & Query2Term-PRF$^*$            & 50.80          & 53.44          & 63.68          & 14.09          & 34.12          & 39.29          & 17.41          & 55.25          & 81.14          & 79.89          & 80.14          & 80.14           \\
                               & Query2Doc$^*$                 & 60.96          & 60.65          & \uline{70.56}  & 17.68          & 41.02          & 47.01          & 22.03          & \textbf{59.88} & \uline{85.25}  & 91.31          & 91.34          & 91.34           \\
                               & Query2Doc-FS$^*$              & 59.95          & \uline{60.67}  & 70.26          & 17.88          & \uline{41.63}  & 47.31          & 21.33          & \uline{59.84}  & 84.52          & 91.33          & 91.37          & 91.37           \\
                               & Query2Doc-PRF$^*$             & \textbf{62.43} & 60.59          & 70.53          & 18.32          & 41.54          & \uline{47.44}  & \textbf{22.35} & 59.63          & 84.93          & 91.42          & 91.45          & 91.45           \\
                               & CoT$^*$                       & 59.90          & 59.15          & 69.35          & 17.16          & 39.59          & 45.50          & 20.57          & 57.50          & 84.22          & \uline{92.19}  & \uline{92.19}  & \uline{92.21}   \\
                               & CoT-PRF$^*$                   & 59.75          & 58.41          & 68.81          & 17.75          & 38.89          & 44.56          & 20.63          & 56.09          & 83.67          & 91.20          & 91.27          & 91.27           \\ 
\cmidrule{2-14}
                               & MILL                          & \uline{61.79}  & \textbf{61.15} & \textbf{71.23} & \textbf{19.05} & \textbf{41.76} & \textbf{48.17} & 21.61          & 59.40          & \textbf{85.27} & \textbf{92.61} & \textbf{92.71} & \textbf{92.72}  \\
\bottomrule
\end{tabular}}
\end{table*}

\subsection{Datasets and Metrics} To evaluate the effectiveness of our proposed method, we conduct extensive experiments on the following public datasets: TREC-DL-2019, TREC-DL-2020, and BEIR. 
\begin{itemize}[leftmargin=*]
    \item \textbf{TREC-DL-2019\&2020 \cite{craswell2021overview}.} TREC-DL-2019 and TREC-DL-2020 \footnote{https://microsoft.github.io/msmarco/} are the datasets used in the TREC Deep Learning Track. We conduct passage retrieval tasks on the datasets, each of which contains 200 queries and 8.84 million passages.
    \item \textbf{BEIR \cite{thakur2021beir}.} BEIR\footnote{https://github.com/beir-cellar/beir} is a heterogeneous benchmark for comprehensive zero-shot evaluation of methods in various information retrieval tasks. We select 9 datasets with small test or dev sets from the 18 available datasets.
\end{itemize}
Following previous work~\cite{claveau2021neural,jagerman2023query,mao2023large}, we use the NDCG@N, MAP@N, Recall@N, and MRR@N as the evaluation metrics, each of which is reported with $\text{N}\in\{10, 100, 1000\}$. Additional experiments on MSMARCO are provided in Appendix~\ref{msmarco}.

\subsection{Baselines}
We conduct comparative experiments with the following baselines, which can be divided into three categories: 
(1) \textbf{Traditional query expansion methods}: Bo1 \cite{amati2002probabilistic}, KL \cite{amati2002probabilistic}, RM3 \cite{abdul2004umass}, and AxiomaticQE \cite{fang2006semantic,yang2019reproducing}. 
(2) \textbf{LLM-based expansion methods}: Query2Term~\cite{jagerman2023query}, Query2Term-FS (the few-shot version of Query2Term), Query2Term-PRF (PRF document augmented Query2Term), Query2Doc~\cite{wang2023query2doc}, Query2Doc-FS, Query2Doc-PRF, CoT~\cite{jagerman2023query}, CoT-PRF. 
(3) \textbf{Ensembled expansion methods}: These are the variants of the LLM-based expansion methods by additionally concatenating top-retrieved PRF documents to the query. They are denoted as Query2Term$^*$, Query2Term-FS$^*$, Query2Term-PRF$^*$, Query2Doc$^*$, Query2Doc-FS$^*$, Query2Doc-PRF$^*$, CoT$^*$, and CoT-PRF$^*$.
The details of the baselines and their prompts are introduced in Appendix~\ref{baselines} and Appendix~\ref{prompts}.

\begin{table*}[!ht]
\centering
\caption{Overall comparison on 9 datasets in BEIR on NDCG@1000. The optimal results are highlighted in bold, while the suboptimal results are underscored. The improvements are all significant (i.e., two-sided t-test with $p<0.05$) between the optimal and suboptimal results.}
\label{tab:beir-NDCG}
\resizebox{\textwidth}{!}{
\begin{tabular}{cccccccccc} 
\toprule
Datasets                      & TREC-COVID     & TOUCHE         & SCIFACT        & NFCORPUS       & DBPEDIA        & FIQA-2018      & SCIDOCS        & ARGUANA        & CLIMATE-FEVER   \\ 
\midrule
No expansion                  & 42.04          & 55.32          & 70.27          & 30.02          & 38.70          & 35.28          & 25.14          & 39.93          & 21.73           \\ 
\midrule
Traditional expansion methods &                &                &                &                &                &                &                &                &                 \\
Bo1                           & 44.73          & 56.62          & 68.34          & 37.01          & 39.05          & 34.97          & 26.14          & 39.42          & 23.11           \\
KL                            & 44.88          & 56.72          & 67.83          & 37.18          & 38.87          & 35.12          & 26.15          & 39.31          & 23.07           \\
RM3                           & 44.54          & 55.79          & 65.28          & 37.27          & 38.11          & 33.14          & 25.91          & 38.14          & 20.71           \\
AxiomaticQE                   & 42.06          & 55.32          & 70.28          & 30.02          & 38.70          & 35.28          & 25.14          & 39.88          & 21.75           \\ 
\midrule
LLM-based expansion methods   &                &                &                &                &                &                &                &                &                 \\
Query2Term                    & 42.48          & 52.95          & 69.57          & 33.82          & 33.51          & 32.12          & 25.11          & 39.33          & 27.23           \\
Query2Term-FS                 & 41.13          & 57.10          & 71.39          & 38.57          & 39.36          & 35.78          & 26.18          & 39.72          & 24.32           \\
Query2Term-PRF                & 39.90          & 53.72          & 60.79          & 38.21          & 34.83          & 31.50          & 24.97          & 38.68          & 23.98           \\
Query2Doc                     & 47.19          & 60.32          & 71.19          & 38.76          & 44.79          & 37.63          & 27.40          & 39.84          & \textbf{32.39}  \\
Query2Doc-FS                  & 46.34          & 59.99          & 71.89          & 38.09          & \uline{45.11}  & 37.96          & 27.18          & 39.92          & \uline{32.05}   \\
Query2Doc-PRF                 & 43.87          & 56.84          & 67.82          & 39.41          & 39.85          & 34.09          & 26.16          & 38.85          & 26.90           \\
CoT                           & 49.32          & 60.77          & 71.63          & 38.88          & 43.05          & 37.28          & \uline{27.50}  & 40.00          & 30.25           \\
CoT-PRF                       & 46.53          & 59.03          & \uline{73.65}  & 39.84          & 40.43          & 38.04          & 26.23          & \uline{40.01}  & 25.78           \\ 
\midrule
Ensembled expansion methods    &                &                &                &                &                &                &                &                &                 \\
Query2Term$^*$                & 48.20          & 59.46          & 68.12          & 41.20          & 39.12          & 35.22          & 26.67          & 39.45          & 26.86           \\
Query2Term-FS$^*$             & 46.54          & 58.03          & 67.96          & 41.12          & 37.78          & 34.60          & 26.01          & 39.58          & 24.97           \\
Query2Term-PRF$^*$            & 45.61          & 56.74          & 64.80          & 39.33          & 36.85          & 33.27          & 25.88          & 39.20          & 23.98           \\
Query2Doc$^*$                 & 50.26          & \uline{61.87} & 71.49          & \uline{41.33}  & 44.06          & 37.05          & 27.49          & 39.49          & 30.67           \\
Query2Doc-FS$^*$              & 50.42          & 62.10          & 72.03          & 41.24          & 44.22          & 37.05          & 27.41          & 39.34          & 29.78           \\
Query2Doc-PRF$^*$             & \uline{50.55}  & 61.74  & 71.86          & 41.20          & 44.47          & 36.82          & 27.47          & 38.49          & 26.25           \\
CoT$^*$                       & 50.69          & 61.69          & 72.33          & 41.08          & 42.29          & \uline{38.13}  & 27.62          & 39.59          & 29.74           \\
CoT-PRF$^*$                   & 47.29          & 59.31          & 70.88          & 40.43          & 38.95          & 36.18          & 26.53          & 39.15          & 25.48           \\ 
\midrule
MILL                          & \textbf{52.53} & \textbf{62.15}          & \textbf{74.14} & \textbf{41.75} & \textbf{46.39} & \textbf{39.23} & \textbf{28.36} & \textbf{40.11} & 30.66           \\
\bottomrule
\end{tabular}}
\end{table*}

\subsection{Implementation Details}
We implement MILL and the baselines with PyTerrier \cite{pyterrier2020ictir}, a Python library helps conduct information retrieval experiments. For the BM25 retriever, we use the default parameters ($b=0.75, k_1=1.2, k_3=8.0$) provided by PyTerrier~\cite{pyterrier2020ictir}. For MILL and all the LLM-based baselines, we use the GPT-3.5-turbo-Instruct API~\cite{brown2020language} provided by OpenAI to generate contextual documents.
The generation parameters are set as $\text{temperature}=0.7$ and $\text{top\_p}=1$. We use the text-embedding-ada-002 provided by OpenAI as the text encoder, where the length of the returned vector is 1536. For other hyperparameters, we set the selection number of generated documents and PRF documents as 3, and the number of candidates as 5. To conduct a fair comparison for the LLM-based baselines, we generate 3 expanded queries for each baseline and concatenate them as the final expansion result. The number of PRF documents for ensembled expansion methods is 3.

\subsection{Main Results}

\begin{table*}
\centering
\caption{Overall comparison on 9 datasets in BEIR on Recall@1000. The optimal results are highlighted in bold, while the suboptimal results are underscored. The improvements are all significant (i.e., two-sided t-test with $p<0.05$) between the optimal and suboptimal results.}
\label{tab:beir-Recall}
\resizebox{\textwidth}{!}{
\begin{tabular}{cccccccccc} 
\toprule
Datasets                      & TREC-COVID     & TOUCHE         & SCIFACT        & NFCORPUS       & DBPEDIA        & FIQA-2018      & SCIDOCS        & ARGUANA        & CLIMATE-FEVER   \\ 
\midrule
No expansion                  & 40.52          & 85.05          & 97.00          & 36.06          & 63.61          & 77.42          & 55.04          & 98.58          & 57.63           \\ 
\midrule
Traditional expansion methods &                &                &                &                &                &                &                &                &                 \\
Bo1                           & 43.64          & \uline{86.00}  & 97.67          & 54.38          & 64.90          & 79.18          & 57.47          & \textbf{98.65} & 60.22           \\
KL                            & 43.63          & \textbf{86.14} & 97.67          & 54.79          & 64.71          & 78.84          & 57.38          & \textbf{98.65} & 60.01           \\
RM3                           & 43.71          & 85.79          & 97.67          & 56.12          & 64.37          & 78.82          & 57.88          & 98.08          & 58.18           \\
AxiomaticQE                   & 40.53          & 85.05          & 97.00          & 36.06          & 63.61          & 77.42          & 55.04          & 98.58          & 57.66           \\ 
\midrule
LLM-based expansion methods   &                &                &                &                &                &                &                &                &                 \\
Query2Term                    & 40.82          & 77.24          & 99.00          & 58.82          & 58.90          & 78.22          & 60.00          & 98.51          & 66.59           \\
Query2Term-FS                 & 40.34          & 85.33          & 98.33          & 61.72          & 65.67          & 81.84          & 60.15          & 98.51          & 62.87           \\
Query2Term-PRF                & 39.50          & 83.29          & 97.50          & 60.55          & 61.11          & 76.31          & 59.25          & \textbf{98.65} & 63.79           \\
Query2Doc                     & 45.42          & 84.08          & 99.00          & 61.09          & \uline{70.29}  & 82.72          & 61.63          & 98.51          & 72.98           \\
Query2Doc-FS                  & 44.66          & 83.95          & 99.33          & 59.55          & 70.04          & 83.46          & 61.33          & 98.36          & \uline{73.01}   \\
Query2Doc-PRF                 & 42.53          & 83.50          & 99.00          & 62.50          & 66.41          & 79.14          & 59.50          & 98.58          & 67.15           \\
CoT                           & 47.27          & 84.42          & 98.67          & 60.63          & 69.24          & \uline{83.56}  & 60.90          & 98.44          & 69.86           \\
CoT-PRF                       & 44.93          & 84.37          & 98.67          & 59.87          & 66.06          & 82.14          & 58.72          & 98.58          & 64.26           \\ 
\midrule
Ensembled expansion methods   &                &                &                &                &                &                &                &                &                 \\
Query2Term$^*$                & 47.04          & 84.92          & 98.67          & 64.66          & 64.77          & 80.36          & 60.82          & 98.44          & 69.09           \\
Query2Term-FS$^*$             & 45.43          & 85.34          & 99.00          & 64.31          & 64.48          & 79.85          & 58.96          & 98.44          & 66.94           \\
Query2Term-PRF$^*$            & 44.67          & 85.42          & 98.83          & 60.80          & 63.37          & 78.84          & 59.61          & 98.44          & 63.79           \\
Query2Doc$^*$                 & 48.43          & 85.49          & 99.33          & 64.23          & 69.95          & 82.47          & 61.32          & \textbf{98.65} & \textbf{73.51}  \\
Query2Doc-FS$^*$              & 48.70          & 85.21          & 99.33          & \uline{64.70}  & 70.23          & 82.40          & 61.05          & \textbf{98.65} & 72.82           \\
Query2Doc-PRF$^*$             & \uline{48.92}  & 84.94          & 99.33          & 63.97          & 70.19          & 82.43          & \uline{61.81}  & 98.58          & 66.42           \\
CoT$^*$                       & 48.87          & 85.16          & 99.33          & 63.82          & 67.63          & 83.32          & 61.69          & 98.51          & 71.64           \\
CoT-PRF$^*$                   & 45.97          & 85.76          & 99.00          & 62.34          & 64.55          & 81.35          & 59.35          & 98.58          & 64.30           \\ 
\midrule
MILL                          & \textbf{50.55} & 85.21          & \textbf{99.67} & \textbf{64.95} & \textbf{71.13} & \textbf{84.23} & \textbf{61.86} & 98.44          & 71.09           \\
\bottomrule
\end{tabular}}
\end{table*}

Tables \ref{tab:overall_result}, \ref{tab:beir-NDCG}, and \ref{tab:beir-Recall} show the experimental results. The full results for the 9 selected datasets in BEIR are listed in Appendix~\ref{beir-results}.
We can draw the following key findings:
\begin{itemize}[leftmargin=*]
    \item Traditional query expansion methods exhibit positive effects for retrieval, while these carefully designed methods are outperformed by Query2Doc and CoT variants by a large margin. This implies that LLM-based methods are more promising for the query expansion task.
    \item Among LLM-based methods, CoT and Query2Doc variants are superior than Query2Term variants. The reason is that generated documents contain more contextualized information than discrete keywords.
    \item Using pseudo-relevance documents and few-shot examples as instructions in LLM-based methods does not necessarily yield positive gains. 
    For instance, Query2Doc-PRF is worse than Query2Doc in TREC-DL-2019 and TREC-DL-2020. This shows that the query expansion task is non-trivial to be aligned to a specific corpus with straightforward prompting techniques.

    \item Ensembled expansion methods (e.g., Query2Doc$^*$) are usually better than LLM-based expansion methods (e.g., Query2Doc), which demonstrates the importance of PRF documents in query expansion. Moreover, MILL is able to outperform the ensembled baselines on most metrics and datasets, as it adopts a more effective combination of generated and PRF documents.

    \item MILL is more effective than all the baselines in general, it always achieves either the best or the second best performance on all metrics and datasets in Table \ref{tab:overall_result} and \ref{tab:beir-NDCG}. It is also worth noting that MILL is a zero-shot method that is more applicable in various real-world applications.
    
\end{itemize}

\subsection{Ablation Study} \label{sec:ablation_study}

We design the following variants of MILL to conduct the ablation study:

\begin{itemize}[leftmargin=*]
    \item \textbf{w/o PRF}: Using QQD to generate expansion directly, without any PRF documents in expansion.
    \item \textbf{w/o MV}:  Concatenating PRF documents to the QQD expansion
    We directly use $K^\prime$ top-retrieved documents of the original query as $\mathcal{D}_{s}^{\text{PRF}}$, 
    without reranking and selection using generated documents $\mathcal{D}^\text{LLM}$.
    \item \textbf{w/o QQD}: Replacing QQD prompt in MILL with Query2Doc prompt.
\end{itemize}

\begin{table}[t]
\caption{Ablation study of MILL on TREC-DL-2020, TREC-COVID and SCIFACT.}
\centering
\label{tab:ablation_study}
\resizebox{0.9\linewidth}{!}{
\begin{tabular}{ccccc} 
\toprule
Methods                          & \begin{tabular}[c]{@{}c@{}}NDCG\\@1000\end{tabular} & \begin{tabular}[c]{@{}c@{}}AP\\@1000\end{tabular} & \begin{tabular}[c]{@{}c@{}}Recall\\@1000\end{tabular} & \begin{tabular}[c]{@{}c@{}}MRR\\@1000\end{tabular}  \\ 
\midrule
\multicolumn{1}{l}{TREC-DL-2020} & \multicolumn{1}{l}{}                                & \multicolumn{1}{l}{}                              & \multicolumn{1}{l}{}                                  & \multicolumn{1}{l}{}                                \\
w/o PRF                          & \uline{70.65}                                       & \uline{48.10}                                     & \textbf{85.97}                                        & 89.10                                               \\
w/o MV                           & 70.28                                               & 46.73                                             & 85.11                                                 & \uline{90.75}                                       \\
w/o QQD                          & 69.46                                               & 47.39                                             & 83.98                                                 & 87.69                                               \\
MILL                             & \textbf{71.23}                                      & \textbf{48.17}                                    & \uline{85.27}                                         & \textbf{92.72}                                      \\ 
\midrule
\multicolumn{1}{l}{TREC-COVID}   & \multicolumn{1}{l}{}                                & \multicolumn{1}{l}{}                              & \multicolumn{1}{l}{}                                  & \multicolumn{1}{l}{}                                \\
w/o PRF                          & 51.17                                               & 27.35                                             & 49.09                                                 & \textbf{92.40}                                      \\
w/o MV                           & \uline{51.73}                                       & \uline{28.44}                                     & \uline{50.00}                                         & 87.40                                               \\
w/o QQD                          & 50.84                                               & 27.30                                             & 49.16                                                 & 89.08                                               \\
MILL                             & \textbf{52.53}                                      & \textbf{29.30}                                    & \textbf{50.55}                                        & \uline{91.17}                                       \\ 
\midrule
\multicolumn{1}{l}{SCIFACT}      & \multicolumn{1}{l}{}                                & \multicolumn{1}{l}{}                              & \multicolumn{1}{l}{}                                  & \multicolumn{1}{l}{}                                \\
w/o PRF                          & \uline{73.01}                                       & \uline{65.58}                                     & \textbf{99.67}                                        & \uline{66.54}                                       \\
w/o MV                           & 72.43                                               & 64.79                                             & \textbf{99.67}                                        & 65.71                                               \\
w/o QQD                          & 71.13                                               & 62.96                                             & \textbf{99.67}                                        & 63.98                                               \\
MILL                             & \textbf{74.14}                                      & \textbf{66.88}                                    & \textbf{99.67}                                        & \textbf{68.09}                                      \\
\bottomrule
\end{tabular}}
\end{table}

Table \ref{tab:ablation_study} shows the results of the ablation study on three datasets, where we can draw the following conclusions: (1) All the three components of MILL have significant contributions to the final performance, (2) \textbf{MILL} is better than \textbf{w/o QQD}, which demonstrates the effectiveness of our proposed QQD prompt. This shows that QQD prompt can effectively leverage the reasoning capabilities of LLMs, assisting LLMs to reveal more diverse and specific search intent, (3) \textbf{MILL} is superior to \textbf{w/o MV}, which verifies the effectiveness of the mutual verification. By mutually selecting the generated and pseudo-relevance documents, it effectively mitigates the corpus unalignment problem of LLMs and compensates for the inaccurate search intent of conventional pseudo-relevance documents, and
(4) Compared to \textbf{w/o PRF} and \textbf{w/o MV}, \textbf{MILL} shows a more significant improvement on BEIR datasets than on TREC-DL-2020. It may indicate that, in specialized domains, mutual verification can more effectively enhance query expansion performance through the use of PRF documents. More results can be found in Appendix~\ref{appendix-ablation}.

\subsection{Varying the Number of Documents}

\begin{figure*}[ht]
    \centering
    \includegraphics[width=\linewidth]{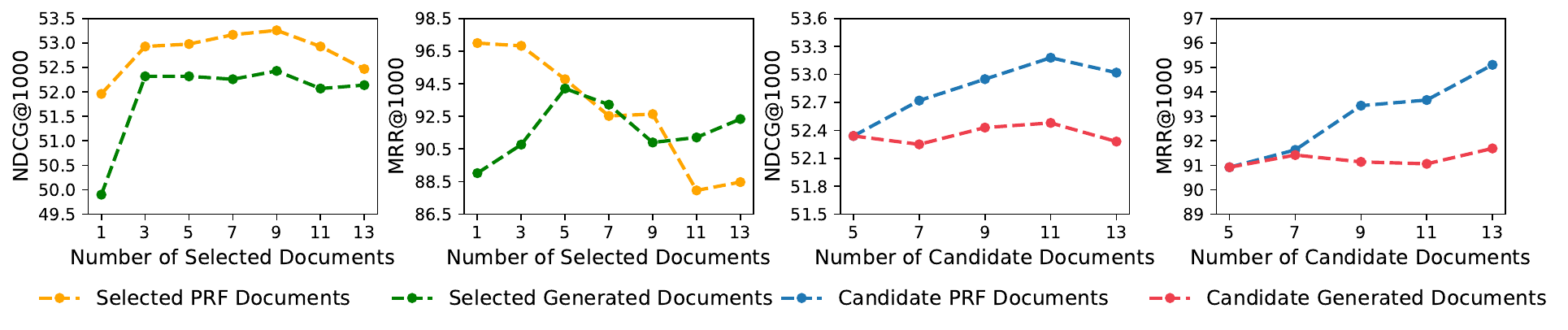}
    \caption{Varying the number of candidate and selected documents on TREC-COVID.}
    \label{fig:hyper}
\end{figure*}

\begin{table*}[!ht]
\centering
\caption{A query expansion example from the TREC-DL-2020 dataset using MILL.}
\label{tab:case_study}
\resizebox{\textwidth}{!}{
\begin{tabular}{p{4cm}p{19cm}} 
\toprule
Query        & What is a nonconformity earth science                                                                                                     \\ 
\midrule
Expansion Content        & \emph{Pseudo-relevance document}: There are three types of unconformities(below): angular \textbf{unconformity}, \textbf{nonconformity}, and disconformity. Anangular \textbf{unconformity} (left) is an erosional \textbf{surface} separating steeply dipping \textbf{rock }layers below from gently dipping layers above. \textbf{Unconformity} is the general term for \textbf{missing} pages of Earth history.                                             \\
             & \emph{LLM-generated document}: 
             \textbf{\textcolor{greyblue}{1.}} What is the definition of a \textbf{\textcolor{greyblue}{nonconformity}} in earth science? A \textbf{\textcolor{greyblue}{nonconformity}} in earth science refers to a type of \textbf{\textcolor{greyblue}{unconformity}}, which is a \textbf{\textcolor{greyblue}{break in the geologic record}} where layers of \textbf{\textcolor{greyblue}{rock}} are \textbf{\textcolor{greyblue}{missing}}. Specifically, a \textbf{\textcolor{greyblue}{nonconformity}} is where \textbf{\textcolor{greyblue}{sedimentary rocks}} are deposited on top of \textbf{\textcolor{greyblue}{igneous or metamorphic rocks}}. 
             \textbf{\textcolor{CamelliaRed}{2.}} How are \textbf{\textcolor{CamelliaRed}{nonconformities}} formed inearth science?~\textbf{\textcolor{CamelliaRed}{Nonconformities}} are formed throughthe process of \textbf{\textcolor{CamelliaRed}{erosion}}, where layers of \textbf{\textcolor{CamelliaRed}{rock}} are gradually wornaway. This exposes the underlying \textbf{\textcolor{CamelliaRed}{igneous or metamorphic rocks}}, which are then \textbf{\textcolor{CamelliaRed}{covered by}} new \textbf{\textcolor{CamelliaRed}{sedimentary rocks}} over time. 
             \textbf{\textcolor{Youhuang}{3.}} What are the different types of \textbf{\textcolor{Youhuang}{nonconformities}} in earth science? There are three main types of \textbf{\textcolor{Youhuang}{nonconformities}}: angular, \textbf{\textcolor{Youhuang}{nonconformity}}, and disconformity. Angular \textbf{\textcolor{Youhuang}{nonconformities occur}} when \textbf{\textcolor{Youhuang}{sedimentary rocks}} are deposited at an angle on top of tilted or folded \textbf{\textcolor{Youhuang}{igneous or metamorphic rocks}}. \textbf{\textcolor{Youhuang}{Nonconformities occur}} when \textbf{\textcolor{Youhuang}{sedimentary rocks}} are deposited on top of \textbf{\textcolor{Youhuang}{eroded igneous ormetamorphic rocks}}. Disconformities \textbf{\textcolor{Youhuang}{occur}} when there is a gap in the \textbf{\textcolor{Youhuang}{sedimentary rock}} layers, indicating \textbf{\textcolor{Youhuang}{a period of erosion}} or non-deposition.
             \textbf{\textcolor{Watergreen}{4.}} What are some examples of \textbf{\textcolor{Watergreen}{nonconformities found}} in the Earth's geology? \textbf{\textcolor{Watergreen}{One well-known example}} of a \textbf{\textcolor{Watergreen}{nonconformity}} is the \textbf{\textcolor{Watergreen}{Grand Canyon}} in Arizona, USA. The \textbf{\textcolor{Watergreen}{sedimentary layers}} seen in the \textbf{\textcolor{Watergreen}{canyon}} were deposited on top of tilted and folded igneous and \textbf{\textcolor{Watergreen}{metamorphic rocks}}, indicating a long history of \textbf{\textcolor{Watergreen}{erosion and deposition}}.  \\ 
\midrule
Ground Truth & \textbf{\textcolor{CamelliaRed}{Nonconformities}} are \textbf{\textcolor{greyblue}{unconformities}} that separate \textbf{\textcolor{greyblue}{igneous or metamorphic rocks}} from overlying \textbf{\textcolor{Youhuang}{sedimentary rocks}}. They usually indicate that along \textbf{\textcolor{Youhuang}{period of erosion occurred}} prior to \textbf{\textcolor{Watergreen}{deposition of the sediments}} (several km of \textbf{\textcolor{CamelliaRed}{erosion}} necessary). They are a feature of stratified \textbf{\textcolor{Watergreen}{rocks}}, and are therefore usually found in \textbf{\textcolor{Youhuang}{sediments}} (but may also occur in stratified volcanics). They are \textbf{surfaces} between two \textbf{rock} bodies that constitute a substantial \textbf{\textcolor{greyblue}{break (hiatus) in the geologic record}} (sometimes people say inaccurately that time is \textbf{missing}). \textbf{Nonconformity}. When \textbf{\textcolor{CamelliaRed}{igneous or metamorphic rocks}} are \textbf{\textcolor{Youhuang}{eroded}} and then \textbf{\textcolor{CamelliaRed}{covered by}} younger \textbf{\textcolor{CamelliaRed}{sedimentary rocks}}, the contact is called a \textbf{nonconformity}. \textbf{\textcolor{Watergreen}{One of the most famous}} of these is \textbf{\textcolor{Watergreen}{found in the Grand Canyon}}, where the oldest \textbf{\textcolor{CamelliaRed}{sedimentary rocks}} are more than a billion years younger than the 1.6 billion-year-old \textbf{\textcolor{Watergreen}{metamorphic rocks}} on which they rest.         \\
\midrule\midrule
Filtered-out PRF document (ranked \#1 by BM25 with the original query) & Definition of nonconformance in the AudioEnglish.org Dictionary. Meaning of non-conformance. What does nonconformance mean? Proper usage of the word nonconformance.  Information about nonconformance in the AudioEnglish.org dictionary, synonyms, and antonyms.                                   \\ 
\midrule
Filtered-out LLM-generated document                                    & ... \textbf{7. }What other geological features are commonly associated with nonconformities in earth science? Nonconformities are often found alongside other geological features, such as faults, folds, and intrusions, which can all provide additional information about the Earth's history and the processes that have shaped it. \textbf{8.} How can nonconformities in earth science be identified in the field? Nonconformities can be identified by looking for the distinct contact between two different rock types, as well as the difference in age between the two layers. Geologists may also use specialized tools, such as radiometric dating, to determine the age of the rocks. \textbf{9.} Are nonconformities only found on land in earth science? No, nonconformities can also be found underwater in the oceans, where layers of sedimentary rock are exposed and show similar  \\
\bottomrule
\end{tabular}}
\end{table*}

In the aforementioned experiments, the default number of candidate (i.e., both generated and retrieved) documents is set to $K=N=5$, and the number of final selected documents is set to $K^\prime=N^\prime=3$. In this subsection, we vary the number of candidates and selected documents and report the performance of MILL on TREC-COVID, w.r.t. NDCG@1000 and MRR@1000. More details and results on additional datasets can be found in Appendix~\ref{appendix-hyper}. 

From Figure~\ref{fig:hyper}, we have observations: 
(1) More selected pseudo-relevance documents decrease MRR@1000 dramatically. This shows that more selected pseudo-relevance documents usually bring more noise to query expansion. In contrast, the generated documents are rather robust, where more selections do not significantly undermine the performance.
(2) When we introduce more candidate documents, the mutual verification framework is able to effectively select pseudo-relevance documents, where both NDCG@1000 and MRR@1000 increase. This shows that LLM-generated documents are very useful for filtering out noisy pseudo-relevance documents. On the other hand, more generated candidate documents do not bring further performance gain, when the number of selected documents is fixed.

\subsection{Case Study}

We show an illustrative example in Table~\ref{tab:case_study}, which contains the original query, the expansion content, and the ground truth (i.e., the human-labeled relevant document). The words of ground truth passage that appear in the pseudo-relevance document are highlighted in bold, and those in the generated documents of different sub-queries are marked with different colors. We can see that the generated document is able to provide more useful information for identifying the ground truth passage. 
We also show the filtered-out PRF document, and the filtered-out LLM-generated document in the table, from which we can observe that the filtered-out documents seem to be 1) PRF documents with limited information and 2) LLM-generated documents with too much extension of the original query. The mutual verification process can filter out these noisy or uninformative documents for MILL.

\section{Related Work}

\textbf{Query Expansion.} Query expansion is a prevalent technique in search platforms, which restructures the original query to more accurately express search intent and enhance 
the alignment with corpus~\cite{bhogal2007review, carpineto2012survey, efthimiadis1996query}. Early studies employed lexical knowledge bases~\cite{qiu1993concept,voorhees1994query} or Pseudo-relevance Feedback (PRF)~\cite{amati2002probabilistic,robertson1990term,rocchio1971relevance,lv2010positional,yan2003multimedia,li2022improving} for expanding the query with additional information. PRF documents can supplement information for any query, but they also encounter the issue of misalignment with the original query~\cite{jagerman2023query}. 

Recently, the integration of LLMs with information retrieval has emerged as a prominent area of research~\cite{li2023agent4ranking,dong2023aligning,sun2023chatgpt,ni2021sentence,zhuang2023rankt5,dai2022promptagator,bonifacio2022inpars,muennighoff2022sgpt}, where LLM-based query expansion methods have also been proposed. In particular, Query2Doc~\cite{wang2023query2doc} proposes a query-document prompt, leveraging the semantic understanding and generative capabilities of LLMs to extend the original query. Another recent study~\cite{jagerman2023query} applies LLMs directly for query expansion across multiple datasets, finding that employing the chain of thoughts (CoT)~\cite{wei2022chain} approach achieves the best results. Moreover, LLMCS~\cite{mao2023large} applies LLMs for query expansion in conversational search, constructing the context search intents as a prompt and combining the chain of thoughts and self-consistency techniques to enhance search performance. 
In our paper, 
we focus on alleviating the limitations of both PRF-based and generation-based method. 
We propose a query-query-document generation method and a mutual verification framework to effective leverage both retrieved and generated contextual documents.

\textbf{Large Language Models.} LLMs have strong and robust abilities in language understanding and generation~\cite{zhao2023survey,kojima2022large,huang2022large,wang2022self}, especially with increased model parameters~\cite{zhao2023survey,jagerman2023query,wei2022emergent}. LLMs have the instruction-following ability~\cite{longpre2023flan,wei2021finetuned} and can be boasted through a few contexts~\cite{min2022rethinking, dong2022survey}, enhancing the performance of LLMs in downstream specific tasks. Moreover, these methods are straightforward and effective, for they require minimal human effort to provide instructions or in-context examples but reach good results. For example, Flan-T5~\cite{chung2022scaling} achieves remarkable results in various NLP downstream tasks by instruction tuning the base model. Recently, many studies~\cite{wei2022chain, besta2023graph, yao2023tree, wang2022self} explored the reasoning capabilities of LLMs and discovered that LLMs are powerful zero-shot reasoners. 
Chain of thoughts~\cite{wei2022chain} (CoT) prompts LLMs to think step by step to activate reasoning capabilities in LLMs.

\section{Conclusion}

In this paper, we propose a novel zero-shot LLMs-based framework for query expansion.
First, we design a QQD prompt scheme that allows LLMs to generate diverse contextual documents via zero-shot reasoning.
Next, we introduce a mutual verification method that allows retrieved and generated contextual documents to complement each other as query expansion.
The experimental results show that our method is superior to the state-of-the-art baselines on three public datasets.

\section*{Acknowledgments}

This research was partially supported by Research Impact Fund (No.R1015-23), APRC - CityU New Research Initiatives (No.9610565, Start-up Grant for New Faculty of CityU), CityU - HKIDS Early Career Research Grant (No.9360163), Hong Kong ITC Innovation and Technology Fund Midstream Research Programme for Universities Project (No.ITS/034/22MS), Hong Kong Environmental and Conservation Fund (No. 88/2022), and SIRG - CityU Strategic Interdisciplinary Research Grant (No.7020046, No.7020074), Ant Group (CCF-Ant Research Fund, Ant Group Research Fund), Huawei (Huawei Innovation Research Program), Tencent (CCF-Tencent Open Fund, Tencent Rhino-Bird Focused Research Program), CCF-BaiChuan-Ebtech Foundation Model Fund, and Kuaishou.

\clearpage
\section{Limitations}

One limitation of our work is the retrieval efficiency. On one hand, during retrieval, MILL needs to perform multiple autoregressive generations for each query based on the query-query-document prompt, and then use mutual verification methods with PRF documents to obtain selected documents. On the other hand, the extended length of the query increases the time required to search the inverted index. To address the issue of multi-round autoregressive generation, $N$ generated documents can be produced in parallel, which will improve generation efficiency. Regarding the issue of extended query length, we can further utilize simple rule-based filtering methods (e.g., deleting words with limited semantic information or truncating documents with word counts) to compress the query.

In addition, from the experiments conducted on the BEIR datasets, we can observe that MILL does not perform well on some metrics for the ARGUANA and CLIMATE-FEVER datasets. This may indicate the limitations of MILL in some scenarios. For ARGUANA, we notice that the queries have 193 words on average, which is roughly 10 to 20 times more words than other BEIR datasets. Thus, it might not necessarily need query expansion, which limits the improvement of MILL. For CLIMATE-FEVER, we observe that
the queries are often declarative sentences, rather than specific questions. In such cases, the QQD approach is more likely to generate off-the-topic subqueries, which undermines the effectiveness of the final query expansion. These observations suggest that MILL could have different performances on different kinds of queries, which will be more comprehensively studied in the future.

\bibliography{reference}
\bibliographystyle{acl_natbib}

\appendix

\clearpage
\appendix
\section{Appendix}

\subsection{Baselines} \label{baselines}

\noindent{\textbf{Traditional query expansion methods}}
\begin{itemize}[leftmargin=*]
    \item \textbf{Bo1 \cite{amati2002probabilistic}.} The Bose-Einstein 1 (Bo1) weighting approach is a method that reconstructs the query based on the frequency of terms found in the feedback documents associated with each query.
    \item \textbf{KL \cite{amati2002probabilistic}.} This method rewrites the queries similar to Bo1 but based on Kullback Leibler divergence.
    \item \textbf{RM3 \cite{abdul2004umass}.} A method used for query expansion in information retrieval, which finds the most relevant terms to the query by using the top-ranked documents returned from the initial query and adds these terms to the original query to create an expanded query.
    \item \textbf{AxiomaticQE \cite{fang2006semantic,yang2019reproducing}.} Axiomatic query expansion (AxiomaticQE) rewrites and expands the origin query by axiomatic semantic term matching.
\end{itemize}
\noindent{\textbf{LLM-based expansion methods}}
\begin{itemize}[leftmargin=*]
    \item \textbf{Query2Term.} It uses LLMs to generate related terms to the origin query in a zero-shot manner. The zero-shot prompts only contain task instructions and the original query.
    \item \textbf{Query2Term-FS.} The few-shot version of Query2Term. The few-shot prompts are built upon zero-shot prompts by adding a few examples. In particular, Query2Term-FS expands upon Query2Term by incorporating additional sets of query-keywords examples.
    \item \textbf{Query2Term-PRF.} It uses the top-3 documents retrieved by the original query as context information to instruct the LLMs to expand the original query.
    \item \textbf{Query2Doc.} The zero-shot version of query2doc \cite{wang2023query2doc}, whose structure is similar to Query2Term. It uses LLMs to generate related passages to the origin query.
    \item \textbf{Query2Doc-FS.} The few-shot version of query2doc \cite{wang2023query2doc}. The prompt structure is similar to Query2Term-FS.
    \item \textbf{Query2Doc-PRF.} It constructs the prompt with pseudo-relevance feedback in a zero-shot manner based on Query2Doc-ZS, like the Query2Term-PRF.
    \item \textbf{CoT.} Chain-of-Thought (CoT) \cite{jagerman2023query} instructs LLMs to generate text step by step, providing a detailed thought process before generating the final answer.
    \item \textbf{CoT-PRF.} A pseudo-relevance feedback based version of CoT similar to Query2Term-PRF.
\end{itemize}
\noindent{\textbf{Ensembled expansion methods}}

The ensembled expansion methods contain Query2Term$^*$, Query2Term-FS$^*$, Query2Term-PRF$^*$, Query2Doc$^*$, Query2Doc-FS$^*$, Query2Doc-PRF$^*$, CoT$^*$, CoT-PRF$^*$. They are the variants to the corresponding LLM-based expansion methods by directly concatenating the top-k PRF documents to the expanded query.

\subsection{Prompts} \label{prompts}

Figure~\ref{tab:q2t_prompt} shows the prompts for the variants of Query2Term. The core prompt is "Write some keywords for the given query: \{query\}."

\begin{table}[h]
\centering
\caption{Prompts for Query2Term and its variants.}
\label{tab:q2t_prompt}
\resizebox{0.95\linewidth}{!}{
\begin{tabular}{ll} 
\toprule
Method         & Prompt    \\ 
\midrule
Query2Term     & Write some keywords for the given query: \textcolor{red}{\{query\}}      \\
\midrule
Query2Term-FS  & \begin{tabular}[c]{@{}l@{}}Write some keywords for the given query:\\\\Context:~\\query:\textcolor{blue}{ \{query1\}}\\keywords:\textcolor[rgb]{0,0.502,0}{ }\textcolor{blue}{\{keywords1\}}\\query:\textcolor[rgb]{0,0.502,0}{ }\textcolor{blue}{\{query2\}}\\keywords: \textcolor{blue}{\{keywords2\}}\\query: \textcolor{blue}{\{query3\}}\\keywords:\textcolor{blue}{ \{keywords3\}}\\\\query: \textcolor{red}{\{query\}}\\keywords:\end{tabular}  \\
\midrule
Query2Term-PRF & \begin{tabular}[c]{@{}l@{}}Write some keywords for the given query:\\\\Context:\\\textcolor{blue}{\{PRF doc 1\}}\\\textcolor{blue}{\{PRF doc 2\}}\\\textcolor{blue}{\{PRF doc 3\}}\\\\query: \textcolor{red}{\{query\}}\\keywords:\end{tabular}                                                                                                                                                                                                        \\
\bottomrule
\end{tabular}}
\end{table}

Figure~\ref{tab:q2d_prompt} shows the prompts for the Query2Doc variants. The main prompts are the sentence: "Write a passage answer the following query: \{query\}."

\begin{table}[h]
\centering
\caption{Prompts for Query2Doc and its variants.}
\label{tab:q2d_prompt}
\resizebox{0.95\linewidth}{!}{
\begin{tabular}{ll} 
\toprule
Method        & Prompt                                                                                                                                                                                                                                                                                                                                                                                                                                             \\ 
\midrule
Query2Doc     & Write a passage answer the following query: \textcolor{red}{\{query\}}                                                                                                                                                                                                                                                                                                                                                                             \\
\midrule
Query2Doc-FS  & \begin{tabular}[c]{@{}l@{}}Write a passage answer the following query:\\\\Context:~\\query:\textcolor{blue}{ \{query1\}}\\passage:\textcolor[rgb]{0,0.502,0}{ }\textcolor{blue}{\{passage1\}}\\query:\textcolor[rgb]{0,0.502,0}{ }\textcolor{blue}{\{query2\}}\\passage: \textcolor{blue}{\{passage2\}}\\query: \textcolor{blue}{\{query3\}}\\passage:\textcolor{blue}{ \{passage3\}}\\\\query: \textcolor{red}{\{query\}}\\passage:\end{tabular}  \\
\midrule
Query2Doc-PRF & \begin{tabular}[c]{@{}l@{}}Write a passage answer the following query:\\\\Context:\\\textcolor{blue}{\{PRF doc 1\}}\\\textcolor{blue}{\{PRF doc 2\}}\\\textcolor{blue}{\{PRF doc 3\}}\\\\query: \textcolor{red}{\{query\}}\\passage:\end{tabular}                                                                                                                                                                                                  \\
\bottomrule
\end{tabular}}
\end{table}

For the CoT and its variants, their prompts are in Figure~\ref{tab:cot_prompt}. The prompts ask LLMs to give the rationale before answering.

\begin{table}[h]
\centering
\caption{Prompts for CoT and its variants.}
\label{tab:cot_prompt}
\resizebox{0.7\linewidth}{!}{
\begin{tabular}{ll} 
\toprule
Method  & Prompt                                                                                                                                                                                                                                                         \\ 
\midrule
CoT     & \begin{tabular}[c]{@{}l@{}}Answer the following query: \textcolor{red}{\{query\}}\\\textcolor{red}{}Give the rationale before answering.\textcolor{red}{}\end{tabular}                                                                                         \\ 
\midrule
CoT-PRF & \begin{tabular}[c]{@{}l@{}}Answer the following query:\\\\Context:\\\textcolor{blue}{\{PRF doc 1\}}\\\textcolor{blue}{\{PRF doc 2\}}\\\textcolor{blue}{\{PRF doc 3\}}\\\\query: \textcolor{red}{\{query\}}\\Give the rationale before answering.\end{tabular}  \\
\bottomrule
\end{tabular}}
\end{table}

\subsection{Results on MSMARCO} \label{msmarco}

Table~\ref{tab:msmarco} shows the experimental results on MSMARCO dataset. MSMARCO\footnote{https://microsoft.github.io/msmarco/}~\cite{nguyen2016ms} is a collection of datasets constructed to advance the development of deep learning in the search field. We choose the passage dataset as our experimental scenario and take \emph{the first 100 queries} from the dev group as the test queries. Results in Table~\ref{tab:msmarco} are based on the LLM text-davinci-003 provided by OpenAI.

\begin{table*}
\centering
\caption{Overall comparison on MSMARCO. The optimal results are highlighted in bold, while the suboptimal results are underscored. The results are reported on NDCG@N, AP@N, Recall@N, and MRR@N with $\text{N}\in\{10, 100, 1000\}$. The improvements are all significant (i.e., two-sided t-test with $p<0.05$) between the optimal and suboptimal results.}
\label{tab:msmarco}
\resizebox{\textwidth}{!}{
\begin{tabular}{ccccccccccccc} 
\toprule
\multirow{2}{*}{Metrics}                          & \multicolumn{3}{c}{NDCG}                                           & \multicolumn{3}{c}{AP}                                             & \multicolumn{3}{c}{Recall}                                         & \multicolumn{3}{c}{MRR}                                             \\ 
\cmidrule{2-13}
                                                  & @10                  & @100                 & @1000                & @10                  & @100                 & @1000                & @10                  & @100                 & @1000                & @10                  & @100                 & @1000                 \\ 
\midrule
No expansion                                      & 28.69                & 34.02                & 36.23                & 23.56                & 24.65                & 24.72                & 44.50                & 69.00                & 86.50                & 22.65                & 23.76                & 23.83                 \\ 
\midrule
\multicolumn{1}{l}{Traditional expansion methods} & \multicolumn{1}{l}{} & \multicolumn{1}{l}{} & \multicolumn{1}{l}{} & \multicolumn{1}{l}{} & \multicolumn{1}{l}{} & \multicolumn{1}{l}{} & \multicolumn{1}{l}{} & \multicolumn{1}{l}{} & \multicolumn{1}{l}{} & \multicolumn{1}{l}{} & \multicolumn{1}{l}{} & \multicolumn{1}{l}{}  \\
Bo1                                               & 29.18                & 33.44                & 35.89                & 23.61                & 24.33                & 24.43                & \uline{46.50}        & 67.50                & 86.50                & \uline{24.07}        & 24.82                & 24.91                 \\
KL                                                & \uline{29.20}        & 33.59                & 36.17                & \uline{23.93}        & \uline{24.73}        & \uline{24.83}        & 45.50                & 66.50                & 86.50                & \textbf{24.39}       & \textbf{25.22}       & \textbf{25.31}        \\
RM3                                               & 26.93                & 32.23                & 34.34                & 21.81                & 22.87                & 22.94                & 42.50                & 67.00                & 83.50                & 22.25                & 23.33                & 23.41                 \\
AxiomaticQE                                       & 28.69                & 34.02                & 36.23                & 23.56                & 24.65                & 24.72                & 44.50                & 69.00                & 86.50                & 22.65                & 23.76                & 23.83                 \\ 
\midrule
\multicolumn{1}{l}{LLM-based expansion methods}   & \multicolumn{1}{l}{} & \multicolumn{1}{l}{} & \multicolumn{1}{l}{} & \multicolumn{1}{l}{} & \multicolumn{1}{l}{} & \multicolumn{1}{l}{} & \multicolumn{1}{l}{} & \multicolumn{1}{l}{} & \multicolumn{1}{l}{} & \multicolumn{1}{l}{} & \multicolumn{1}{l}{} & \multicolumn{1}{l}{}  \\
Query2Term                                        & 23.28                & 29.50                & 32.00                & 19.74                & 21.01                & 21.08                & 34.17                & 63.17                & 83.67                & 19.91                & 21.17                & 21.24                 \\
Query2Term-FS                                     & 24.26                & 29.76                & 32.07                & 20.41                & 21.43                & 21.50                & 36.33                & 62.50                & 81.33                & 20.78                & 21.87                & 21.94                 \\
Query2Term-PRF                                    & 21.56                & 27.02                & 29.26                & 16.04                & 17.05                & 17.12                & 38.67                & 64.83                & 83.33                & 16.04                & 17.11                & 17.17                 \\
Query2Doc                                         & 25.83                & 31.31                & 33.82                & 20.27                & 21.33                & 21.42                & 43.50                & 69.00                & 88.83                & 20.39                & 21.50                & 21.58                 \\
Query2Doc-FS                                      & 28.23                & 33.22                & 35.89                & 23.10                & 23.99                & 24.09                & 44.67                & 68.83                & \uline{89.50}        & 23.00                & 23.94                & 24.04                 \\
Query2Doc-PRF                                     & 25.45                & 29.99                & 32.36                & 20.31                & 21.25                & 21.33                & 41.44                & 62.50                & 81.17                & 20.45                & 21.35                & 21.43                 \\
CoT                                               & 26.13                & 31.84                & 34.25                & 21.38                & 22.44                & 22.54                & 41.00                & 68.33                & 86.83                & 21.47                & 22.55                & 22.64                 \\
CoT-PRF                                           & 28.93                & \uline{34.17}        & \uline{36.32}        & 23.51                & 24.52                & 24.60                & 46.12                & \uline{70.87}        & 87.50                & 23.64                & 24.69                & 24.77                 \\
MILL                                              & \textbf{29.99}       & \textbf{34.92}       & \textbf{37.26}       & \textbf{24.01}       & \textbf{24.98}       & \textbf{25.07}       & \textbf{48.67}       & \textbf{71.67}       & \textbf{89.83}       & 24.02                & \uline{25.02}        & \uline{25.10}         \\
\bottomrule
\end{tabular}}
\end{table*}

\subsection{More Results on BEIR} \label{beir-results}

In this section, we list the full results for the 9 selected datasets from BEIR. Specifically, they are TREC-COVID, TOUCHE, SCIFACT, NFCORPUS, DBPEDIA, FIQA-2018, SCIDOCS, ARGUANA, and CLIMATE-FEVER. The optimal results are highlighted in bold, while the suboptimal results are underscored. The results are reported on NDCG@N, AP@N, Recall@N, and MRR@N with N (10, 100, 1000)

\begin{table*}
\centering
\caption{Overall experimental results on TREC-COVID.}
\label{tab:trec-covid}
\resizebox{\textwidth}{!}{
\begin{tabular}{ccccccccccccc} 
\toprule
\multirow{2}{*}{Metrics}      & \multicolumn{3}{c}{NDCG}                         & \multicolumn{3}{c}{AP}                          & \multicolumn{3}{c}{Recall}                      & \multicolumn{3}{c}{MRR}                           \\ 
\cmidrule{2-13}
                              & @10            & @100           & @1000          & @10           & @100           & @1000          & @10           & @100           & @1000          & @10            & @100           & @1000           \\ 
\midrule
No expansion                  & 62.59          & 47.41          & 42.04          & 1.46          & 8.16           & 19.79          & 1.74          & 11.91          & 40.52          & 83.37          & 83.37          & 83.37           \\ 
\midrule
Traditional expansion methods &                &                &                &               &                &                &               &                &                &                &                &                 \\
Bo1                           & 64.82          & 49.5           & 44.73          & 1.56          & 8.8            & 22.01          & 1.77          & 12.48          & 43.64          & 86.62          & 86.77          & 86.77           \\
KL                            & 65.8           & 49.93          & 44.88          & 1.59          & 8.9            & 22.26          & 1.79          & 12.51          & 43.63          & 86.62          & 86.79          & 86.79           \\
RM3                           & 64.05          & 48.5           & 44.54          & 1.55          & 8.62           & 21.87          & 1.78          & 11.22          & 43.71          & 82.96          & 83.06          & 83.06           \\
AxiomaticQE                   & 62.74          & 47.45          & 42.06          & 1.47          & 8.17           & 19.81          & 1.74          & 11.91          & 40.53          & 84.37          & 84.37          & 84.37           \\
\bottomrule
\end{tabular}}
\end{table*}

\begin{table*}
\centering
\caption{Overall experimental results on TOUCHE.}
\label{tab:touche}
\resizebox{\textwidth}{!}{
\begin{tabular}{ccccccccccccc} 
\toprule
\multirow{2}{*}{Metrics}      & \multicolumn{3}{c}{NDCG}                         & \multicolumn{3}{c}{AP}                           & \multicolumn{3}{c}{Recall}                       & \multicolumn{3}{c}{MRR}                           \\ 
\cmidrule{2-13}
                              & @10            & @100           & @1000          & @10            & @100           & @1000          & @10            & @100           & @1000          & @10            & @100           & @1000           \\ 
\midrule
No expansion                  & 34.28          & 45.48          & 55.32          & 13.06          & 20.96          & 22.47          & 20.69          & 54.92          & 85.05          & 62.28          & 62.71          & 62.71           \\ 
\midrule
Traditional expansion methods &                &                &                &                &                &                &                &                &                &                &                &                 \\
Bo1                           & 35.62          & 46.98          & 56.62          & 14.19          & 22.19          & 23.69          & 21.35          & 56.47          & \uline{86.00}  & 63.54          & 64.07          & 64.07           \\
KL                            & 35.52          & 46.96          & 56.72          & 14.00          & 22.18          & 23.68          & 20.99          & 56.78          & \textbf{86.14} & 63.98          & 64.51          & 64.51           \\
RM3                           & 34.66          & 46.54          & 55.79          & 13.72          & 22.00          & 23.42          & 22.03          & 57.79          & 85.79          & 56.73          & 57.09          & 57.09           \\
AxiomaticQE                   & 34.28          & 45.48          & 55.32          & 13.06          & 20.96          & 22.47          & 20.69          & 54.92          & 85.05          & 62.28          & 62.71          & 62.71           \\
\bottomrule
\end{tabular}}
\end{table*}

\begin{table*}
\centering
\caption{Overall experimental results on SCIFACT.}
\label{tab:scifact}
\resizebox{\textwidth}{!}{
\begin{tabular}{ccccccccccccc} 
\toprule
\multirow{2}{*}{Metrics}      & \multicolumn{3}{c}{NDCG}                         & \multicolumn{3}{c}{AP}                           & \multicolumn{3}{c}{Recall}                       & \multicolumn{3}{c}{MRR}                           \\ 
\cmidrule{2-13}
                              & @10            & @100           & @1000          & @10            & @100           & @1000          & @10            & @100           & @1000          & @10            & @100           & @1000           \\ 
\midrule
No expansion                  & 67.22          & 69.66          & 70.27          & 62.11          & 62.67          & 62.7           & 81.43          & 92.27          & 97             & 63.24          & 63.66          & 63.68           \\ 
\midrule
Traditional expansion methods &                &                &                &                &                &                &                &                &                &                &                &                 \\
Bo1                           & 65.14          & 67.63          & 68.34          & 59.3           & 59.92          & 59.95          & 81.59          & 92.2           & 97.67          & 60.42          & 60.87          & 60.89           \\
KL                            & 64.68          & 67.08          & 67.83          & 58.69          & 59.28          & 59.31          & 81.59          & 91.87          & 97.67          & 59.76          & 60.18          & 60.21           \\
RM3                           & 62.22          & 64.54          & 65.28          & 55.45          & 55.97          & 55.99          & 81.34          & 91.93          & 97.67          & 56.24          & 56.58          & 56.61           \\
AxiomaticQE                   & 67.22          & 69.66          & 70.28          & 62.11          & 62.68          & 62.7           & 81.43          & 92.27          & 97             & 63.24          & 63.66          & 63.68           \\
\bottomrule
\end{tabular}}
\end{table*}

\begin{table*}
\centering
\caption{Overall experimental results on NFCORPUS.}
\label{tab:nfcorpus}
\resizebox{\textwidth}{!}{
\begin{tabular}{ccccccccccccc} 
\toprule
\multirow{2}{*}{Metrics}      & \multicolumn{3}{c}{NDCG}                         & \multicolumn{3}{c}{AP}                           & \multicolumn{3}{c}{Recall}                       & \multicolumn{3}{c}{MRR}                           \\ 
\cmidrule{2-13}
                              & @10            & @100           & @1000          & @10            & @100           & @1000          & @10            & @100           & @1000          & @10            & @100           & @1000           \\ 
\midrule
No expansion                  & 32.22          & 27.29          & 30.02          & 12.08          & 14.36          & 14.89          & 14.78          & 24.38          & 36.06          & 53.44          & 53.82          & 53.83           \\ 
\midrule
Traditional expansion methods &                &                &                &                &                &                &                &                &                &                &                &                 \\
Bo1                           & 33.49          & 30.21          & 37.01          & 12.73          & 15.98          & 17.09          & 16.26          & 29.71          & 54.38          & 52.74          & 53.24          & 53.28           \\
KL                            & 33.56          & 30.22          & 37.18          & 12.73          & 15.89          & 17.01          & 16.3           & 29.61          & 54.79          & 53.49          & 53.99          & 54.03           \\
RM3                           & 33.41          & 30.31          & 37.27          & 12.36          & 15.68          & 16.8           & 16.82          & 30.46          & 56.12          & 52.35          & 52.81          & 52.85           \\
AxiomaticQE                   & 32.22          & 27.29          & 30.02          & 12.08          & 14.36          & 14.89          & 14.78          & 24.38          & 36.06          & 53.44          & 53.82          & 53.83           \\
\bottomrule
\end{tabular}}
\end{table*}

\begin{table*}
\centering
\caption{Overall experimental results on DBPEDIA.}
\label{tab:dbpedia}
\resizebox{\textwidth}{!}{
\begin{tabular}{ccccccccccccc} 
\toprule
\multirow{2}{*}{Metrics}      & \multicolumn{3}{c}{NDCG}                         & \multicolumn{3}{c}{AP}                           & \multicolumn{3}{c}{Recall}                       & \multicolumn{3}{c}{MRR}                           \\ 
\cmidrule{2-13}
                              & @10            & @100           & @1000          & @10            & @100           & @1000          & @10            & @100           & @1000          & @10            & @100           & @1000           \\ 
\midrule
No expansion                  & 26.59          & 32.45          & 38.7           & 11.59          & 17.71          & 18.89          & 17.2           & 42.15          & 63.61          & 51.7           & 52.37          & 52.39           \\ 
\midrule
Traditional expansion methods &                &                &                &                &                &                &                &                &                &                &                &                 \\
Bo1                           & 26.59          & 32.59          & 39.05          & 11.65          & 18.03          & 19.24          & 17.32          & 42.67          & 64.9           & 50.47          & 51.17          & 51.2            \\
KL                            & 26.42          & 32.44          & 38.87          & 11.52          & 17.89          & 19.09          & 17.27          & 42.62          & 64.71          & 50.01          & 50.84          & 50.86           \\
RM3                           & 25.47          & 31.81          & 38.11          & 10.88          & 17.4           & 18.6           & 17.05          & 42.92          & 64.37          & 46.6           & 47.28          & 47.31           \\
AxiomaticQE                   & 26.59          & 32.45          & 38.7           & 11.59          & 17.71          & 18.89          & 17.2           & 42.15          & 63.61          & 51.7           & 52.37          & 52.39           \\
\bottomrule
\end{tabular}}
\end{table*}

\begin{table*}
\centering
\caption{Overall experimental results on FIQA-2018.}
\label{tab:fiqa-2018}
\resizebox{\textwidth}{!}{
\begin{tabular}{ccccccccccccc} 
\toprule
\multirow{2}{*}{Metrics}      & \multicolumn{3}{c}{NDCG}                         & \multicolumn{3}{c}{AP}                          & \multicolumn{3}{c}{Recall}                       & \multicolumn{3}{c}{MRR}                           \\ 
\cmidrule{2-13}
                              & @10            & @100           & @1000          & @10            & @100           & @1000         & @10            & @100           & @1000          & @10            & @100           & @1000           \\ 
\midrule
No expansion                  & 25.26          & 31.74          & 35.28          & 19.4           & 20.86          & 21.04         & 30.97          & 55.92          & 77.42          & 31.03          & 32.11          & 32.18           \\ 
\midrule
Traditional expansion methods &                &                &                &                &                &               &                &                &                &                &                &                 \\
Bo1                           & 24.36          & 31.21          & 34.97          & 18.71          & 20.3           & 20.49         & 30.21          & 56.25          & 79.18          & 29.37          & 30.51          & 30.58           \\
KL                            & 24.75          & 31.4           & 35.12          & 18.99          & 20.52          & 20.72         & 30.88          & 56.21          & 78.84          & 29.77          & 30.84          & 30.92           \\
RM3                           & 22.8           & 29.23          & 33.14          & 16.85          & 18.32          & 18.51         & 30.37          & 54.82          & 78.82          & 26.47          & 27.55          & 27.63           \\
AxiomaticQE                   & 25.26          & 31.76          & 35.28          & 19.4           & 20.87          & 21.04         & 30.97          & 56             & 77.42          & 31.03          & 32.11          & 32.18           \\
\bottomrule
\end{tabular}}
\end{table*}

\begin{table*}
\centering
\caption{Overall experimental results on SCIDOCS.}
\label{tab:scidocs}
\resizebox{\textwidth}{!}{
\begin{tabular}{ccccccccccccc} 
\toprule
\multirow{2}{*}{Metrics}      & \multicolumn{3}{c}{NDCG}                         & \multicolumn{3}{c}{AP}                         & \multicolumn{3}{c}{Recall}                       & \multicolumn{3}{c}{MRR}                          \\ 
\cmidrule{2-13}
                              & @10            & @100           & @1000          & @10          & @100           & @1000          & @10            & @100           & @1000          & @10           & @100           & @1000           \\ 
\midrule
No expansion                  & 14.71          & 20.91          & 25.14          & 8.36         & 9.73           & 9.94           & 15.84          & 34.48          & 55.04          & 25.37         & 26.41          & 26.48           \\ 
\midrule
Traditional expansion methods &                &                &                &              &                &                &                &                &                &               &                &                 \\
Bo1                           & 15.1           & 21.82          & 26.14          & 8.73         & 10.29          & 10.51          & 16.43          & 36.39          & 57.47          & 25.31         & 26.41          & 26.48           \\
KL                            & 15.1           & 21.81          & 26.15          & 8.75         & 10.31          & 10.54          & 16.37          & 36.24          & 57.38          & 25.43         & 26.54          & 26.61           \\
RM3                           & 14.56          & 21.49          & 25.91          & 8.41         & 10.05          & 10.28          & 15.79          & 36.24          & 57.88          & 24.46         & 25.63          & 25.7            \\
AxiomaticQE                   & 14.71          & 20.91          & 25.14          & 8.36         & 9.73           & 9.94           & 15.84          & 34.48          & 55.04          & 25.37         & 26.41          & 26.48           \\
\bottomrule
\end{tabular}}
\end{table*}

\begin{table*}
\centering
\caption{Overall experimental results on ARGUANA.}
\label{tab:arguana}
\resizebox{\textwidth}{!}{
\begin{tabular}{ccccccccccccc} 
\toprule
\multirow{2}{*}{Metrics}      & \multicolumn{3}{c}{NDCG}                         & \multicolumn{3}{c}{AP}                           & \multicolumn{3}{c}{Recall}              & \multicolumn{3}{c}{MRR}                           \\ 
\cmidrule{2-13}
                              & @10            & @100           & @1000          & @10            & @100           & @1000          & @10            & @100           & @1000 & @10            & @100           & @1000           \\ 
\midrule
No expansion                  & \textbf{34.24} & 39.36          & 39.93          & 22.55          & 23.70          & 23.72          & \textbf{71.27} & 94.17          & 98.58 & 22.56          & 23.71          & 23.73           \\ 
\midrule
Traditional expansion methods &                &                &                &                &                &                &                &                &       &                &                &                 \\
Bo1                           & 33.07          & 38.92          & 39.42          & 21.81          & 23.14          & 23.16          & 68.78          & \textbf{94.74} & \textbf{98.65} & 21.82          & 23.15          & 23.17           \\
KL                            & 32.90          & 38.79          & 39.31          & 21.66          & 23.01          & 23.03          & 68.49          & 94.59          & \textbf{98.65} & 21.68          & 23.02          & 23.04           \\
RM3                           & 30.81          & 37.29          & 38.14          & 20.75          & 22.20          & 22.23          & 62.52          & 91.61          & 98.08 & 20.76          & 22.21          & 22.24           \\
AxiomaticQE                   & 34.19          & 39.30          & 39.88          & 22.49          & 23.64          & 23.66          & \textbf{71.27} & 94.10          & 98.58 & 22.50          & 23.65          & 23.67           \\
\bottomrule
\end{tabular}}
\end{table*}

\begin{table*}
\centering
\caption{Overall experimental results on CLIMATE-FEVER.}
\label{tab:climate-fever}
\resizebox{\textwidth}{!}{
\begin{tabular}{ccccccccccccc} 
\toprule
\multirow{2}{*}{Metrics}      & \multicolumn{3}{c}{NDCG}                         & \multicolumn{3}{c}{AP}                           & \multicolumn{3}{c}{Recall}                       & \multicolumn{3}{c}{MRR}                           \\ 
\cmidrule{2-13}
                              & @10            & @100           & @1000          & @10            & @100           & @1000          & @10            & @100           & @1000          & @10            & @100           & @1000           \\ 
\midrule
No expansion                  & 12.52          & 17.96          & 21.73          & 8.28           & 9.32           & 9.48           & 16.20          & 35.80          & 57.63          & 17.07          & 18.26          & 18.35           \\ 
\midrule
Traditional expansion methods &                &                &                &                &                &                &                &                &                &                &                &                 \\
Bo1                           & 13.48          & 19.43          & 23.11          & 8.86           & 10.03          & 10.20          & 17.65          & 39.10          & 60.22          & 18.14          & 19.32          & 19.40           \\
KL                            & 13.48          & 19.42          & 23.07          & 8.90           & 10.06          & 10.23          & 17.50          & 38.92          & 60.01          & 18.25          & 19.43          & 19.51           \\
RM3                           & 11.24          & 17.17          & 20.71          & 6.80           & 7.99           & 8.15           & 16.39          & 37.64          & 58.18          & 13.99          & 15.21          & 15.28           \\
AxiomaticQE                   & 12.54          & 17.98          & 21.75          & 8.28           & 9.33           & 9.49           & 16.23          & 35.82          & 57.66          & 17.09          & 18.28          & 18.37           \\
\bottomrule
\end{tabular}}
\end{table*}

\subsection{More Results for Ablation Experimetns} \label{appendix-ablation}
From the results shown in Table~\ref{table:appendix-ablation}, we can draw some findings that MILL consistently achieves better performance than w/o PRF, w/o MV, and w/o QQD on all three datasets. This validates the effectiveness of both QQD and mutual verification across different datasets.

\begin{table*}
\caption{More results of ablation experiments.}
\label{table:appendix-ablation}
\centering
\resizebox{\textwidth}{!}{
\begin{tabular}{ccccccccccccc} 
\toprule
\multirow{2}{*}{Methods}         & \multicolumn{3}{c}{NDCG}                                           & \multicolumn{3}{c}{AP}                                             & \multicolumn{3}{c}{Recall}                                         & \multicolumn{3}{c}{MRR}                                             \\
                                 & @10                  & @100                 & @1000                & @10                  & @100                 & @1000                & @10                  & @100                 & @1000                & @10                  & @100                 & @1000                 \\ 
\midrule
\multicolumn{1}{l}{TREC-DL-2020} & \multicolumn{1}{l}{} & \multicolumn{1}{l}{} & \multicolumn{1}{l}{} & \multicolumn{1}{l}{} & \multicolumn{1}{l}{} & \multicolumn{1}{l}{} & \multicolumn{1}{l}{} & \multicolumn{1}{l}{} & \multicolumn{1}{l}{} & \multicolumn{1}{l}{} & \multicolumn{1}{l}{} & \multicolumn{1}{l}{}  \\
w/o PRF                          & 60.13                & 59.88                & \uline{70.65}        & \textbf{19.63}       & \uline{41.57}        & \uline{48.10}        & \textbf{22.47}       & 58.34                & \textbf{85.97}       & 88.97                & 89.09                & 89.10                 \\
w/o MV                           & 59.46                & 59.42                & 70.28                & 18.01                & 40.18                & 46.73                & 21.45                & \uline{58.66}        & 85.11                & \uline{90.70}        & \uline{90.75}        & \uline{90.75}         \\
w/o QQD                          & \uline{60.26}        & \uline{59.89}        & 69.46                & 17.96                & 41.56                & 47.39                & 21.19                & 58.55                & 83.98                & 87.65                & 87.69                & 87.69                 \\
\bottomrule
\end{tabular}}
\end{table*}

\subsection{More Results for Experiments with Various Numbers of Documents} \label{appendix-hyper}

In this subsection, we will supplement the results on other metrics for the experiments with various numbers of documents. We use the gpt-3.5-turbo-instruct API provided by OpenAI to conduct these experiments.

\begin{figure*}
    \centering
    \includegraphics[width=\linewidth]{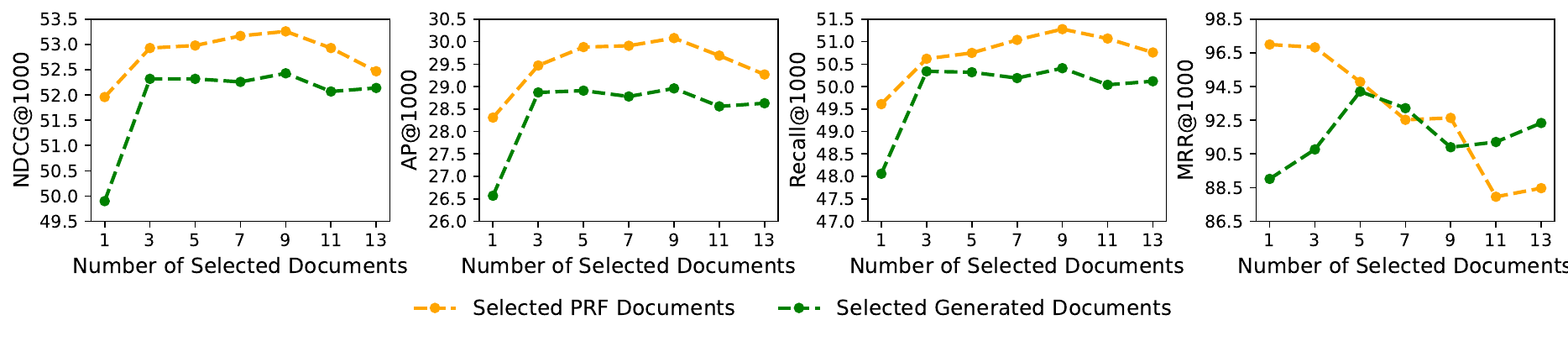}
    \caption{Hyperparameter analysis on the number of document selections on TREC-COVID. The x-axis denotes the number of documents selected, and the y-axis represents the metrics values (NDCG@1000, AP@1000, Recall@1000, and MRR@1000).}
    \label{fig:hyper-1}
\end{figure*}

\begin{figure*}
    \centering
    \includegraphics[width=\linewidth]{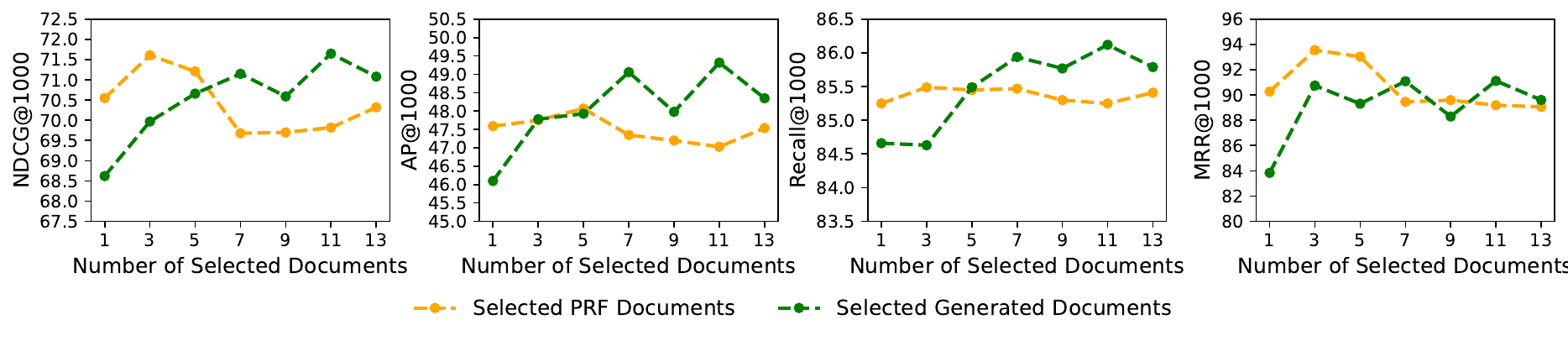}
    \caption{Hyperparameter analysis on the number of document selections on TREC-DL-2020. The x-axis denotes the number of documents selected, and the y-axis represents the metrics values (NDCG@1000, AP@1000, Recall@1000, and MRR@1000).}
    \label{fig:hyper-1-dl-2020}
\end{figure*}

The experiments concerning the number of selected documents are shown in Figure~\ref{fig:hyper-1} and Figure~\ref{fig:hyper-1-dl-2020}. When the number of selected generated documents changes, the number of candidate generated documents remains 15, and the number of PRF candidate documents and the number of selected PRF documents remain 5 and 3. When the number of selected PRF documents changes, the number of candidate PRF documents remains 15, and the number of generated candidate documents and the number of selected generated documents remain 5 and 3.
We can find that the trends of selected PRF documents in NDCG, AP, and Recall are consistent, yet contrary to that of MRR. This is due to the fact that NDCG, AP, and Recall are more comprehensive indicators, whereas MRR only considers the ranking of the topmost relevant document retrieved.

\begin{figure*}
    \centering
    \includegraphics[width=\linewidth]{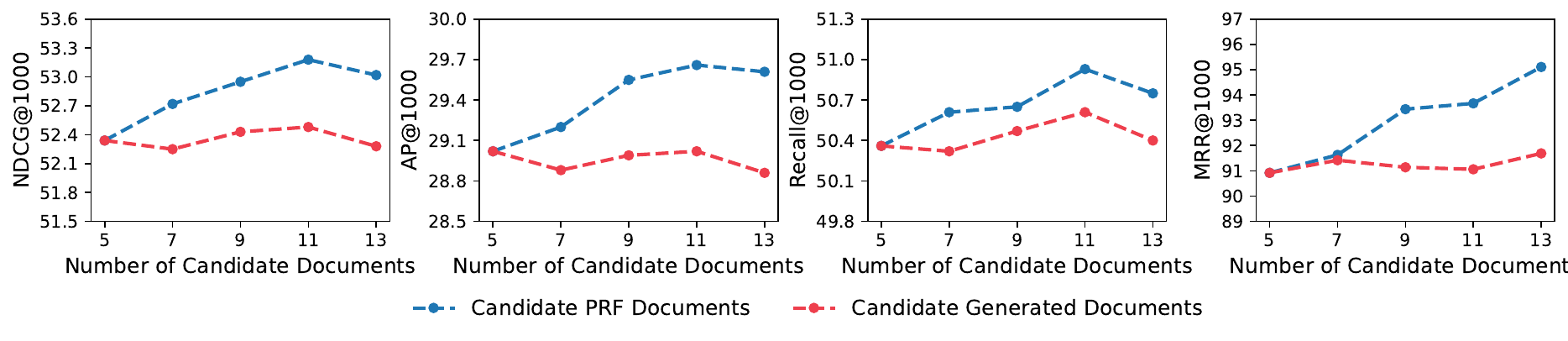}
    \caption{Hyperparameter analysis on the number of document candidates on TREC-COVID. The x-axis denotes the number of document candidates, and the y-axis represents the metrics values (NDCG@1000, AP@1000, Recall@1000, and MRR@1000).}
    \label{fig:hyper-2}
\end{figure*}

\begin{figure*}
    \centering
    \includegraphics[width=\linewidth]{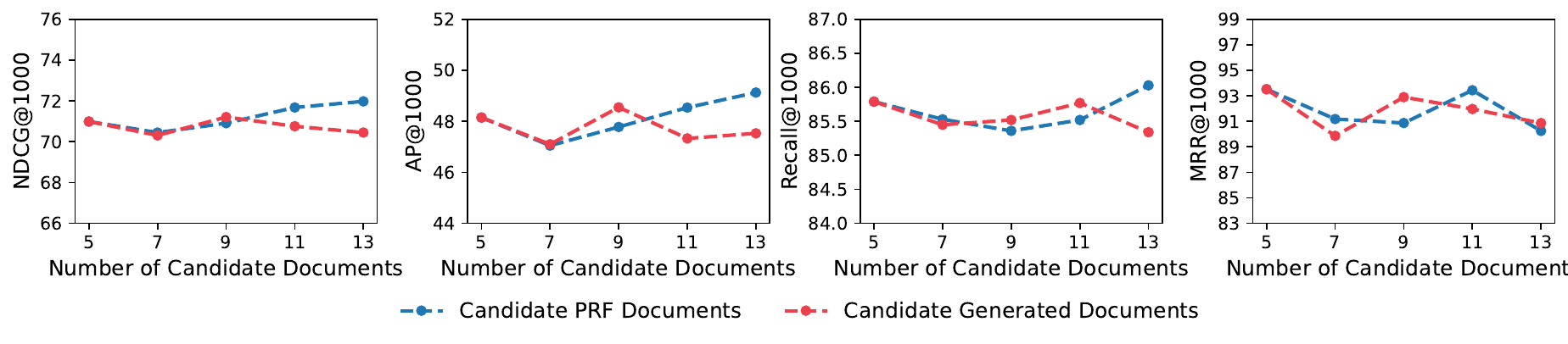}
    \caption{Hyperparameter analysis on the number of document candidates on TREC-DL-2020. The x-axis denotes the number of document candidates, and the y-axis represents the metrics values (NDCG@1000, AP@1000, Recall@1000, and MRR@1000).}
    \label{fig:hyper-2-dl-2020}
\end{figure*}

In the experiments regarding the number of candidate documents, as shown in Figure~\ref{fig:hyper-2} and Figure~\ref{fig:hyper-2-dl-2020}, we can observe a similar trend across different metrics: as the number of generated document candidates increases, the metrics remain relatively stable. However, with an increase in the number of PRF document candidates, there is a noticeable growth in the metrics. This suggests that a specific number of generated documents, such as 5, can almost entirely cover the additional information provided by the generation process to aid in understanding the search intent of the original query. Meanwhile, PRF documents, derived from searches based on the original query, suggest that more PRF document candidates can cover a wider range of possible search intents, thereby enhancing the effectiveness of query expansion.

\end{document}